\documentclass[letterpaper,twoside,12pt]{article}
\usepackage{graphicx,natbib,amsmath,amssymb}
\title{X-ray Emission From Accretion Disks of AGN:
Signatures of Supermassive Black Holes}

\author{Predrag Jovanovi\'{c} and Luka \v{C}. Popovi\'{c} \\
Astronomical Observatory, Volgina 7, 11160 Belgrade, Serbia}

\date{}

\begin{document}

\maketitle

\tableofcontents
\markright{CONTENTS}

\vfil\newpage

\section*{Abstract}
\markright{Abstract}
\addcontentsline{toc}{section}{Abstract}

In this chapter we discuss the X-ray radiation from relativistic
accretion disks around supermassive black holes, supposed to exist
in the centers of Active Galactic Nuclei (AGN). Our focus is on the
X-ray radiation, especially in the Fe K$\alpha$ line which
originates in the innermost parts of an accretion disk. Moreover,
here we discuss some effects which can disturb the Fe K$\alpha$
profile and cause its rapid and irregular variability, observed in
the X-ray spectra of some AGN. We will pay attention to three such
effects: perturbations in the disk emissivity, absorbtion by warm
absorbers and gravitational microlensing. The X-ray emission from
accretion disks around non-rotating (Schwarzschild metric), as well
as rotating (Kerr metric) supermassive black holes, is discussed.
The X-ray radiation of AGN is probably produced in a compact region
near their central supermassive black holes, and can provide us some
essential information about the plasma conditions and the space-time
geometry in these regions. The goal of this chapter is mainly to
present a short overview of some important and recent investigations
in this field.

\section{Introduction}
\markright{Introduction}

Active Galactic Nuclei are powerful sources of radiation in a wide
spectral range: from $\gamma$ rays to radio waves. They derive their
extraordinary luminosities (sometimes more than $10.000$ times
higher than luminosities of "ordinary" galaxies) from energy release
by matter accreting towards, and falling into, a central
supermassive black hole \citep[see e.g.][]{Peterson04}.

According to the unification model of AGN \citep{Antonucci93}, their
central engine consists of a supermassive black hole (SMBH) with
mass ranging from $10^5$ to $10^9$ solar masses $M_\odot$
\citep{Kaspi00,Peterson04b}, which is surrounded by an accretion
disk that radiates in the X-ray band \citep{Fabian89}. The vast
majority of the X-ray sources in the Universe are AGN. The integral
emission of AGN reflects the history of accretion onto SMBH over
cosmic time. Emission lines are usually seen in the X-ray spectra of
AGN. The broad emission Fe K$\alpha$ spectral line (6.4--6.9~keV,
depending on ionization state) with asymmetric profile (narrow
bright blue peak and a wide faint red wing) has been observed in a
number of type 1 AGN \citep[see e.g.][]{Nandra07}. Early results
from the ASCA (\emph{Advanced Satellite for Cosmology and
Astrophysics}) era suggested that broad relativistic lines might be
common in type 1 AGN. However, surprisingly, they have been detected
in significant number and their features described in only a small
fraction of those sources \citep{Reynolds03}. The first and the best
studied one is MCG-6-30-15 \citep{Tanaka95,Fabian03}.

In some cases the line width corresponds to one third of speed of
light, indicating that its emitters rotate with relativistic
velocities. Therefore, the line is probably produced in a very
compact region near the central black hole of AGN and can provide us
some essential information about the plasma conditions and the
space-time geometry in vicinity of the black hole. Consequently, if
the line is emitted close enough to the SMBH, it shows a broad
relativistic profile affected by SMBH spin \citep{Reynolds08} and
gravitational redshift (as well as other general relativistic
effects) \citep{Fabian89,Laor91}.

Several studies have been performed over samples of local AGN
\citep[see
e.g.][etc]{Nandra97,Yaqoob05,Nandra07,Bianchi08,Markowitz08}, as
well as from distant quasars \citep{Corral08} in order to
characterize the Fe K$\alpha$ emission. \citet{Nandra07} performed a
spectral analysis of a sample of 26 type 1 to 1.9 Seyferts galaxies
($z < 0.05$) observed by \emph{XMM-Newton}. They found that a
relativistic line is significantly detected in a half of their
sample (54$\pm$10 percent) with a mean equivalent width (EW) of
$\sim$~80~eV, but around 30\% of selected AGN showed a relativistic
broad line that can be explained by the emission of an accretion
disk.

Accretion disks could have different forms, dimensions, and
emission, depending on the type of central black hole, whether it is
rotating (Kerr metric) or non-rotating (Schwarzschild metric). They
represent an efficient mechanism for extracting gravitational
potential energy and converting it into radiation, giving us the
most probable explanation for the main characteristics of AGN (high
luminosity, compactness, jet formation, rapid time variations in
radiation and the profile of the Fe K$\alpha$ spectral line).

Here we discuss the X-ray radiation from relativistic accretion
disks around supermassive black holes, supposed to exist in the
centers of AGN. Especially, we discuss the Fe K$\alpha$ line profile
which originates from the accretion disk. Moreover, we also present
some effects which can disturb the Fe K$\alpha$ profile, such as:
perturbations in the disk emissivity, absorbtion by warm absorbers
and gravitational microlensing.

The aim of this chapter is to present a short overview of results of
some recent investigations in this field, and it is divided into the
following six sections: \emph{Active Galactic Nuclei as Hosts of
Supermassive Black Holes} - where the main features, classification
and unified model of active galaxies are briefly presented,
\emph{Space-Time Geometry in Vicinity of Supermassive Black Holes} -
where the basic definitions of Schwarzschild and Kerr metrics are
given, \emph{Accretion Disk Around a Supermassive Black Hole} -
where we explain the standard model of an accretion disk, including
its emission, accretion rate, luminosity, structure and spectral
distribution, \emph{Supermassive Black Holes and X-ray Emission} -
where the focus is on the modeling of the observed X-ray radiation
from relativistic accretion disk around a supermassive black hole of
AGN in both the Fe K$\alpha$ spectral line and X-ray continuum,
\emph{Variability of X-ray Emission Around Supermassive Black Hole}
- where we present the main causes of rapid and irregular
variability of the X-ray emission which can be due to disk
instability, reflecting in perturbations of its emissivity, or it
could be caused by some external effects, such as gravitational
microlensing and absorption by X-ray absorbers, and finally
\emph{Conclusion} - where the most important results from previous
sections are pointed out and their brief summary is given.

Finally, we should note that a huge number of papers devoted to
investigation of the X-ray emission of AGN was published in the last
two decades, so it was not possible to mention all of them here, but
more details and references can be found in review papers such as
e.g. \citet{Brandt05,Miller07,Harris06}, etc.

\section{Active Galactic Nuclei as Hosts of Su\-per\-mas\-si\-ve Black Ho\-les}
\markright{Active Galactic Nuclei as Hosts of Supermassive Black Holes}

Active galaxies differ from the so called "normal" galaxies in the
amount of energy emitted from their nuclei. The term \emph{Active
Galactic Nuclei} (AGN) refers to the energetic phenomena in the
central region of a galaxy which cannot be solely or directly
produced by stars. The nebular-like emission spectra of NGC 1068,
NGC 4051 and NGC 4151, with H$\alpha$, [O II] $\lambda$3727, [Ne
III] $\lambda$3869 and [O III] $\lambda\lambda$ 4363,4959,5007 were
observed by \citet{Hubble26}, but the first classification of these
objects was made by \citet{Seyfert43}. He recognized these objects
as a class of galaxies with strong and high-excitation optical
emission lines localized in the nuclei\footnote{For more detailed
historical review see \citet{Osterbrock89}}.

Active galaxies have been the subject of an intensive astrophysical
investigations for the last 3 -- 4 decades. As the most luminous
objects with one of the most powerful energy release rates and with
the most compact dimensions, AGN are of the great interest in modern
astrophysics. AGN (i.e. quasars) are the brightest objects in the
Universe and thus, these objects are important for the studies of
the early Universe and cosmology in general. For instance, studies
of their luminosity functions or quasar host galaxies are crucial
for understanding the formation and evolution of galaxies in
general. The class of active galaxies contains many different
objects, such as quasars, gigantic elliptic radio galaxies, luminous
spiral Seyfert galaxies, blazars, etc.

\subsection{Main characteristics of AGN}

AGN are the most luminous objects in the Universe and they have
luminosities in the range from $\sim 10^{42}$ to $10^{48}$ erg
s$^{-1}$, that can be $10^4$ times greater than in the case of a
typical galaxy. Apart from their great luminosities, AGN are
emitting in the broad band of electromagnetic spectrum. The observed
emission of AGN is in the continuum and in lines, from the $\gamma$
and X domains to the far infrared and radio bands \citep[see
e.g.][]{Peterson04}.

One of the characteristics of AGN is violent and fast variability
observed in different parts of electromagnetic spectra. Fast changes
in the AGN brightness are observed sometimes just for a couple of
days, e.g. in the case of galaxy NGC 4151 \citep{Shapovalova08}.
This leads to the conclusion that the emission regions in the AGN
have small dimensions (from only couple of light days up to couple
of light months). Also, the brightness of AGN can be so high that
the brightness of the host galaxy can be neglected in the total
brightness (for instance in the case of quasars). Consequently, they
are compact regions which could not be directly observed in the most
bands of electromagnetic spectrum. Therefore, AGN and their
physical/kinematical properties are indirectly studied through the
analysis of their spectra.

In contrast to the spectral energy distribution (SED) of an ordinary
galaxy (that represents a sum of stellar spectra, thus the most of
its luminosity comes within no more than one decade of frequency),
AGN emit across the broad range of frequencies (of order $10^5$ Hz).
The AGN continuum in a number of high-luminosity sources peaks in
the ultraviolet, i.e. in the so called Big Blue Bump (BBB)
\citep{Shilds78,Zhou97}. The BBB is attributed to the thermal
emission from an optically thick region of the accretion disk
\citep{Shilds78,Ulrich80}, or from optically thin regions, i.e.
free-free emission \citep{Antonucci88,Ferland90}. The observed SED
in AGN is remarkably different from the thermal (black-body)
spectrum of a star or a regular galaxy \citep{Oke68}. This
featureless continuum observed in AGN characterizes the main source
of energy input and suggests that the mechanisms that produce it are
common to all types of AGN. The SED can be approximated with the
power law function $F\approx\nu^\alpha$ where $\alpha$ is the
spectral index \citep{Krolik99,Peterson04}.

The observed continuum of a typical Seyfert 1 galaxy is very strong
compared with Seyfert 2 galaxies; as a result Seyfert 1 galaxies
appear to be more luminous than Seyfert 2-s. Some AGN have the SED
almost flat from the infrared to X-ray part of the spectrum, so that
the spectral index is $\alpha\sim 1$, although it is usually
steeper. In the radio band, the brightness of active galaxies is
usually an order of magnitude higher than of normal galaxies. But
still, even if it is so intensive, the radio luminosity is never
higher more than 1\% of the bolometric luminosity.

Another property of AGN is strong X-ray flux, with the X-ray
continuum produced by lower energy photons which are Compton
scattered to higher energies by relativistic electrons
\citep{Sunyaev80}. The fraction of the power emitted in the X-ray
emission is three or four times larger in AGN than in normal
galaxies \citep{Krolik99,Peterson04}, therefore the X-emission can
indicate the presence of AGN.

Today, it is widely accepted that the mechanism which powers AGN is
accretion onto supermassive black hole (SMBH), i.e. the greatest
part of the continuum emission comes from an accretion disk and its
corona. The mass of the supermassive black hole is estimated to be
from $10^5$ to $10^9\ M_\odot$ \citep{Peterson04b,Peterson00}. Thus,
in the center of an AGN there is a supermassive black hole with an
accretion disk surrounded by gas and dust in the form of a torus.

There are also other emitting regions present in the vicinity of the
supermassive black hole, such as ionized gas clouds that produce
intensive emission lines or jets of matter that are mostly visible
in the radio band, but sometimes they can also be seen in the
optical band. The kinematics of these regions is very complex as
well as the physical processes that produce such specific spectra of
these objects. In the central part, there are mainly three emitting
line regions: Fe K$\alpha$ (in the X-ray range)\footnote{See \S 5.2
for more details}, Broad and Narrow Line Emitting Regions (BLR and
NLR, respectively; in optical/UV spectra).

In the optical and UV band, the total flux of emission (and
occasionally absorbtion) lines contributes from several percent to
tens of percent of the continuum flux. The existence of the broad
and narrow emission lines comes from the fact that an AGN contains
two separated emission line regions with different kinematics,
density, ionization, optical thickness and radiation transfer.

The basic characteristics of the BLR and NLR can be described as
follows:
\begin{list}{-}{}
\item BLR is a compact region, with dimensions from only couple of
    light days up to couple of light weeks \citep{Kaspi00,Peterson04b}, that is located in the vicinity of the
    black hole ($< $1 pc). The structure of this region is very
    complex and most likely it consists from at least two separated
    subregions (see e.g.\citet{Popovic04,Ilic06,Bon06,Popovic08}, etc).
    The ionized gas in this region is of relatively high density
    ($n > 10^9 \mathrm{cm^{-3}}$) and temperature $T\sim 10^4$ K.
    Emitters are moving with high velocities, up to 10000 km/s.
    In this region the broad emission lines (BELs) from allowed transitions are formed.
\item NLR extends even up to 1 kpc far from the central source
    \citep{Peterson04}. The density in this region is significantly
    smaller ($n \sim 10^3\rm cm^{-3}$) than in the BLR, while the temperature is on
    the same order of magnitude as in the BLR. From this region the
    narrow emission lines (NELs) arise, indicating a random motion of the
    emitting gas smaller than 1000 km$\rm s^{-1}$. NELs are often coming
    from the forbidden transitions. The physical properties of this
    region are much closer to the properties of emission nebulae than to the BLR.
\end{list}

According to the different characteristics mentioned above, there
are several classes of AGN.

\subsection{Classification}

The first group of stellar systems with active nuclei consists of
Seyfert galaxies. These are spiral galaxies with very bright nuclei
which spectra have strong emission lines of neutral and
multi-ionized emitters. The presence of highly ionized emission
lines indicates the existence of a non-stellar ionization continuum.
The presence or absence of BELs (i.e. BLR) has been historically
used to separate Seyfert galaxies into two classes. Seyfert 1
galaxies have broad permitted (H I, He I and He II) and narrow
permitted and forbidden lines (such as [O III] $\lambda$5007), while
Seyfert 2 galaxies have only narrow permitted and forbidden lines
\citep{Khachikian74}.  This simple Seyfert classification scheme can
be further sub-divided according to specific spectral properties
\citep{Osterbrock89}, as e.g. Seyfert galaxies with intermediate
Balmer profiles are classified as Seyfert 1.5-s with apparent narrow
Balmer line components superimposed on broad wings.

Many low luminosity galaxies have a nucleus that resembles the
Seyfert 2 nucleus, but with the spectrum that shows the forbidden
lines from lower ionization states. Therefore, these galaxies are
called Low-Ionization Nuclear Emission-line Regions, i.e LINERs.

Active galaxies which emit intensively in the radio band are so
called radio galaxies (RG). Radio galaxies are dominantly elliptic
galaxies that are divided into two morphologically different types:
FR I and FR II. The classification is done according to the relative
position of the surface brightness maximum (called the hot spot) in
the radio lobes. For example, in the FR I radio galaxy the distance
between the hot spot in the radio lobes is less than half of the
maximum diameter of the radio source, while in the FR II type this
distance is larger \citep{Binney98}. Like Seyfert galaxies, radio
galaxies can be divided into subgroups according to the widths of
their emission lines, i.e. Narrow Line Radio Galaxies (NLRG) that
emit narrow emission lines characteristic for Seyfert 2 galaxies,
and Broad Line Radio Galaxies (BLRG) that emit broad lines as
Seyfert 1 galaxies.

Being the most powerful sources in the Universe, quasars belong to
the group of active galaxies with the absolute magnitude $M_B < -23$
\citep{Osterbrock89}. These quasi stellar (quasars) objects emit
intensively in the radio domain and their spectrum shows broad
emission lines significantly redshifted. In this group of AGN there
are also Quasi-Stellar Objects (QSOs) which have the same observed
properties as quasars, except they are weak radio sources. Besides,
there is a group of QSOs with broad absorption lines, so called
Broad Absorption Line Quasars (BAL QSOs).

Finally, the separate class of AGN are so called BL Lacerte (Lacs)
objects. BL Lacs are nuclei of elliptical galaxies that emit highly
variable and polarized radiation, and have a non thermal optical
continuum and a strong radio emission. With this group of galaxies,
we often connect another type of AGN, Optically Violently Variable
(OVVs) quasars. The OVVs, in contrast to BL Lacs which have
featureless continuum, emit broad optical emission lines
characteristic for quasars. Both these groups of objects are often
referred to as one class called blazars.

\subsection{Unified model}

As it was mentioned above, all types of AGN have some common
properties. It is widely accepted that central engine of an AGN is
accretion onto supermassive black hole. First observational evidence
in favor of a unified model of AGN was  spectropolarometric
observations of Seyfert 2 galaxy NGC 1068 \citep{Antonucci85}.
\citet{Antonucci85} found broad Balmer lines and Fe II emission,
characteristic of a Seyfert 1 spectrum, in the polarized spectrum of
NGC 1068.

\begin{figure}[ht!]
\centering
\includegraphics[width=0.9\textwidth]{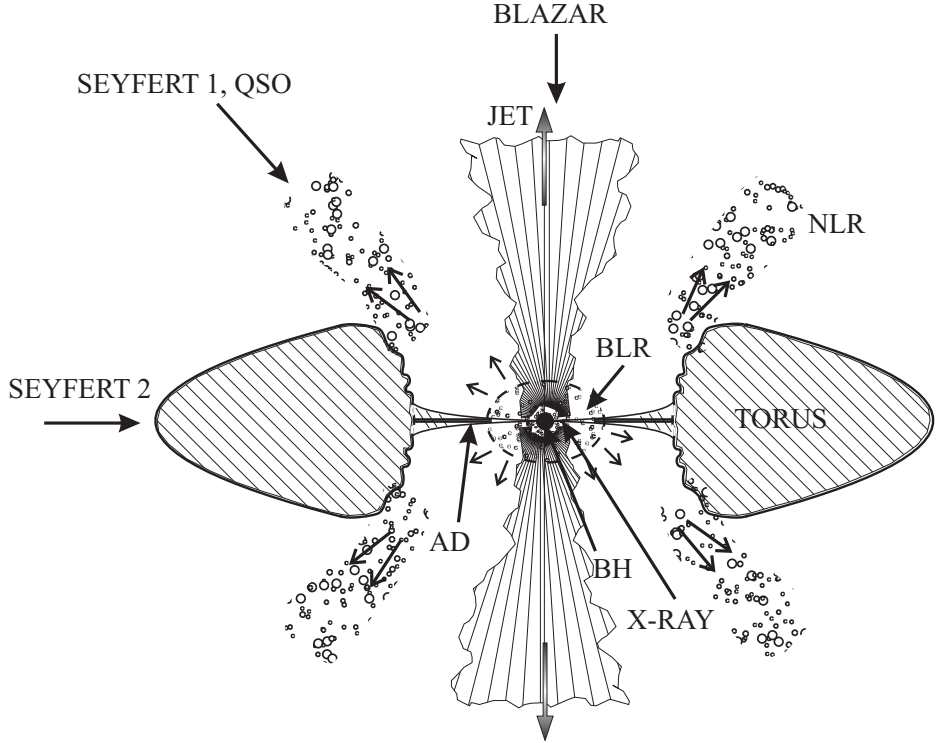}
\caption{The scheme of unified model of AGN (Figure courtesy: Vesna Borka Jovanovi\' c).}
\label{fig23_1}
\end{figure}

In the unified model of AGN \citep{Antonucci93} both types of
Seyfert galaxies are intrinsically the same. The difference in
spectral characteristics appears due to different angle of the
visibility of the central regions (see Fig. \ref{fig23_1}).

A geometrically and optically thick dusty molecular torus-like
structure surrounds the central source, as well as the BLR (see Fig.
1). Therefore the visibility of the nuclear engine depends on
viewing angle with respect to the torus. The broad-line polarization
in Seyfert 2-s is probably due to electron scattering. This picture
is consistent with the differences observed between continuum fluxes
of the Seyfert 1 and Seyfert 2 galaxies \citep{Clavel00}.

Following the unified model, the line-of-sights to Seyfert 2
galaxies (or Narrow Line Objects) are obstructed by optically thick
material corresponding to hydrogen column densities of $N_H >
10^{22}\ \mathrm{cm^{-2}}$ \citep{Risaliti99}. For column densities
$N_H < 10^{24}\ \mathrm{cm^{-2}}$, photons above a few keV can
penetrate the torus creating an un-obstructed view of the nuclear
source. In this case the source is called \emph{Compton thin}. For
column densities $10^{25}>N_H>10^{24}\ \mathrm{cm^{-2}}$,
 only high energy X-ray emission (tens of keV) can pass through
the obscuring material \citep{Turner97}. But in the case $N_H >
10^{25}\ \mathrm{cm^{-2}}$ even high energy X-rays, above a few tens
of keV, are Compton scattered and the nuclear source is completely
hidden from our direct view. Therefore, by observing near the
edge-on torus, one can detect narrow line objects, with weak (or
without) X-ray emission; while by observing near the face-on torus,
one can detect Blazars or OVVs. The broad line objects can be
detected if one observes an inclined torus.

\section{Space-Time Geometry in Vicinity of Su\-per\-mas\-si\-ve Black Ho\-les}
\markright{Space-Time Geometry in Vicinity of Supermassive Black Holes}

As it was shown in the previous section, according to unified model
of AGN, in their heart there is a supermassive black hole with mass
up to $10^9\ M_\odot$, surrounded by an accretion disk. A black hole
is a region of space-time around some collapsed mass which
gravitational field became so powerful that nothing (including
electromagnetic radiation) could escape from its attraction, after
crossing the certain boundary called the event horizon
\citep{Hawking88}.

All black holes in nature are commonly classified according to their
mass as: supermassive black holes (with masses $M_{BH}\sim
10^5-10^9\ M_\odot$), intermediate-mass black holes ($M_{BH}\sim
10^2-10^5\ M_\odot$), stellar-mass black holes ($M_{BH} < 10^2\
M_\odot$), mini and micro black holes ($M_{BH} \ll M_\odot$). In
this chapter our focus will be on supermassive black holes located
in the centers of AGN, because they are surrounded by accretion
disks which X-ray radiation is the main subject of this discussion.
Space-time geometry in vicinity of a supermassive black hole depends
on its type (more precisely, on its angular momentum), being either
non-rotating, where the local space-time geometry is described by
Schwarzschild metric, or rotating, where the Kerr metric determines
the geometry of local space-time.

\subsection{Historical background}

The term black hole was introduced in 1969 by the American scientist
John Wheeler, but the predecessor of the modern idea of black holes
emerged in the late eighteenth century. It was independently
developed by the British geologist John Michell in 1783 and the
French mathematician and astronomer Pierre-Simon Laplace in 1796.
This idea was based on Newton's theory of gravity and particle
theory of light, and is commonly referred to as a dark star. A dark
star is, according to Michell, a sufficiently massive and compact
star which have such a strong gravitational field that the
corresponding surface escape velocity equals or exceeds the speed of
light. Therefore, any light emitted from the surface of such star
would be dragged back by the gravitational field and we would not be
able to see it, but we could still feel its gravitational attraction
\citep{Hawking88}. After the wave theory of light was founded, the
idea about a dark star was forgotten since, according to the wave
theory, it was not clear that light could be affected by gravity at
all.

Modern theory of black holes was founded in the twentieth century,
more precisely in 1915, shortly after Einstein introduced the theory
of general relativity, when the German physicist and astronomer Karl
Schwarzschild found a particular exact solution to the Einstein
field equations, for the limited case of a single spherical
non-rotating mass.

The implications of the general relativity to the stellar evolution
were firstly understood by Subrahmanyan Chandrasekhar in 1928
\citep{Hawking88}. He calculated a maximum limit for the mass of a
star, above which the star would not be able to support itself (by
repulsion of electron degeneracy pressure which arises from Pauli
exclusion principle) against its own gravity, after it had used up
all its fuel. This limit ($\approx 1.5\ M_\odot$) is nowadays known
as the Chandrasekhar limit. If a star's mass is less than this
limit, it can eventually stop contracting at a final state known as
a "white dwarf" \citep{Hawking88}. A similar discovery was made in
1932 by the Russian scientist Lev Davidovich Landau. He found that
there is another limiting mass ($\approx 1 - 2\ M_\odot$), for which
the corresponding final state is much smaller than a white dwarf and
which would be supported against its gravity by the exclusion
principle repulsion between neutrons and protons, rather than
between electrons. Such stars are therefore called neutron stars
\citep{Hawking88}. All these results led to a surprising conclusion
that a star which mass is above the Chandrasekhar limit, when it
comes to the end of its fuel, could experience a catastrophic
gravitational collapse to a point with an infinite density. American
scientist Robert Oppenheimer was the first who explained in 1939
that, according to general relativity, stars above $\approx 3\
M_\odot$ would at the end of their lives collapse into black holes,
due to reasons given by Chandrasekhar \citep{Hawking88}.

With the discovery of QSOs in the early 60s, the speculations about
how their enormous luminosities could be produced within such a
small region occurred. Thermonuclear reactions were not enough
efficient, and therefore, they were quickly eliminated. Hoyle and
Fowler suggested in 1963 that such luminosities could be produced
during the collapses and disintegrations of stelar-type objects with
masses of $10^5 - 10^8$ $M_{\odot}$ \citep{Ferrarese05}. Soon, it
became clear that the energy source was gravitational. For example,
Wheeler suggested a mechanism in which gravitational singularity at
the center of a galaxy converted the falling matter into energy
\citep{Ferrarese05}. Zel'dovich, Novikov and Salpeter proposed in
1964 the release of energy through accretion due the growth of a
massive object at the center of a galaxy, while Lynden-Bell in 1969
made an attempt to explain the phenomena observed in QSOs and
Seyfert galaxies in terms of a black hole formalism
\citep{Ferrarese05}. A crucial event for the acceptance of black
holes was the discovery of pulsars by Jocelyn Bell in 1967, because
it was the clear evidence of the existence of neutron stars, and
therefore, confirmation of Chandrasekhar limit. The first detection
of a solar mass black hole came in 1972, when the mass of the
rapidly variable X-ray source Cygnus X-1 was proven to be above the
maximum allowed for a neutron star \citep{Ferrarese05}. The first
convincing evidence for the existence of supermassive black holes is
the discovery of the Fe $K\alpha$ spectral line in X-ray spectrum of
Seyfert galaxy MGC-6-30-15 by \citet{Tanaka95}.\footnote{See \S 5.2
for more details}

Nowadays, the black hole theory is very well established, mainly due
to the contributions of numerous authors during the second half of
the twentieth century, like Roy Kerr, Werner Israel, John Wheeler,
Brandon Carter, Roger Penrose, Stephen Hawking and many
others.\footnote{For more detailed historical overview on this
topic, see e.g. \citet{Hawking88}}

\subsection{Schwarzschild metric - non-rotating black hole}

Schwarzschild metric describes the space-time geometry in
spherically symmetric gravitational field around a time-steady
non-rotating black hole in vacuum. The solution of the Einstein
field equations under these conditions was found by Karl
Schwarzschild in 1916 \citep{Schwarzschild16a,Schwarzschild16b}. If
we denote the Schwarzschild radius which corresponds to a mass $M$
by $R_S=\dfrac{2GM}{c^2}\approx 3\dfrac{M}{M_\odot}\mathrm{km}$
(where $G$ is the gravitational constant and $c$ is the speed of
light), then the squared element of space-time interval $ds^2$ in
the case of Schwarzschild metric is given by:
\begin{equation}
ds^2=\left(\dfrac{1}{1-\dfrac{R_S}{r}}\right)dr^2
-\left(1-\dfrac{R_S}{r}\right)c^2dt^2
+r^2(d\theta^2+\sin^2{\theta}d\phi^2),
\label{eq31_1}
\end{equation}
where $t$ is the time coordinate measured by a stationary clock at
infinity (which must be distinguished from the proper time $\tau$,
measured by a clock moving with the particle), $r$ is the radial
coordinate, $\theta$ is the colatitude (angle from North) in radians
and $\phi$ is the longitude in radians.

It is obvious from Eq. (\ref{eq31_1}) that Schwarzschild metric
depends only on the mass $M$ through the Schwarzschild radius $R_S$.
The Schwarzschild metric has an infinite space-time curvature (i.e.
a singularity) at $r = 0$. In the case when $r\rightarrow\infty$
this metric is reduced to the Minkowski metric of flat space-time:
$ds^2=dr^2-c^2dt^2+r^2(d\theta^2+\sin^2{\theta}d\phi^2)$.

If the complete mass $M$ collapses below its Schwarzschild radius
$R_S$, then it will form a spherically symmetric Schwarzschild black
hole. A Schwarzschild black hole has no charge or angular momentum
and it can be distinguished from any other Schwarzschild black hole
only by its mass.

To a first approximation, the matter surrounding a black hole may be
assumed to rotate in circular Keplerian orbits with velocity
\citep[e.g.][]{Shakura73}: $v^2=\dfrac{GM}{R}$. Therefore, taking
into account that a maximum allowed rotation velocity is speed of
light ($v=c$), one can obtain so called gravitational radius $R_g$:
\begin{equation}
R_g=\dfrac{GM}{c^2}=\dfrac{R_S}{2}\approx 1.5\dfrac{M}{M_\odot}\mathrm{km},
\label{eq31_2}
\end{equation}
which is usually used as a unit for distance around a black hole.

\subsection{Kerr metric - rotating black hole}

Roy Kerr, a New Zealand mathematician, found an exact solution of
the Einstein field equations of general relativity in the case of
gravitational field outside an uncharged rotating black hole. The
properties of Kerr solution, known as Kerr metric, are given in
numerous papers and books \citep[see
e.g.][]{Chandrasekhar83,Carter73} and therefore, here we will
present only its definition and some basic features, necessary for
the further discussion.

Contrary to the Schwarzschild case, the Kerr metric is no longer
spherically symmetric. In the case of a rotating black hole with
mass $M$ and angular momentum $J=a M$, this metric is given by
\citep[see e.g.][]{Fanton97}:
\begin{equation}
ds^2=-\left(1-{\dfrac{2Mr}{\Sigma}}\right)dt^2
-{\dfrac{4Mar}{\Sigma}}\sin^2{\theta}dt d\phi
+{\dfrac{A}{\Sigma}}\sin^2{\theta}d\phi^2
+{\dfrac{\Sigma}{\Delta}}dr^2+\Sigma d\theta^2,
\label{eq32_1}
\end{equation}
where
\begin{equation}
\Sigma=r^2+a^2\cos^2{\theta}, \hspace*{0.5cm} \Delta=r^2+a^2-2Mr, \hspace*{0.5cm}
A=(r^2+a^2)^2-a^2\Delta\sin^2{\theta}.
\label{eq32_2}
\end{equation}
Above definition is given in Boyer-Lindquist coordinates for
$c=G=1$. As it can be seen from expressions (\ref{eq32_1}) and
(\ref{eq32_2}), in addition to black hole mass $M$, the Kerr metric
also depends on its specific angular momentum $a$. Here $a\leq M$
because any additional angular momentum would increase the energy of
the black hole and, therefore, its mass \citep{Krolik99}. In the
case of a non-rotating black hole, i.e. for $a\rightarrow 0$, it
reduces to the Schwarzschild metric.

By solving the equation $\Delta=0$ and taking only a root with $+$
sign, we obtain the radius of event horizon of a black hole:
$r_h=M+\sqrt{M^2-a^2}.$ Thus, when $a\rightarrow 0$ then
$r_h\rightarrow 2M=R_S$ (because we assumed that $c=G=1$), whereas
when $a\rightarrow M$ (i.e. for a maximally rotating black hole)
then $r_h\rightarrow M$.

The minimum allowed radius $r_{ms}$ of a stable circular equatorial
orbit, or so called marginally stable orbit, is given by the roots
of the equation \citep{Fanton97}: $r^2-6Mr\mp 8a\sqrt{Mr}-3a^2=0$,
where the upper sign refers to co-rotating orbits, while the lower
one to counter-rotating orbits. In the case of Schwarzschild metric
($a=0$) $r_{ms}=6M$, whereas for a  maximally rotating black hole
($a=M$) $r_{ms}=M$ \citep{Fanton97}.

Rotating black holes are surrounded by a region called the
ergosphere, in which there can be no static observers and the
negative energy orbits are possible \citep{Krolik99}. Also, there is
a general relativistic effect known as frame-dragging, according to
which a rotating black hole drags the space-time around itself, but
such effects will not be discussed here in more details.

\section{Accretion Disk Around a Supermassive $ $ Black Hole}
\markright{Accretion Disk Around a Supermassive Black Hole}

Some observed quasars have luminosities of up to $L_{bol} \sim
10^{47}$ erg/s \citep{Schneider06}. The corresponding total energy
emitted during the lifetime of such quasars can be estimated to $E
\geq 3\times 10^{61}$ erg, assuming that their minimum age is about
$10^7$ years and that their luminosities  do not change
significantly over the lifetime \citep{Schneider06}. Only two known
mechanisms could produce such enormous amount of energy: nuclear
fusion and accretion. The maximum efficiency $\epsilon$ (defined as
the mass fraction of fuel that is converted into energy, according
to $E=\epsilon m c^2$) of thermonuclear reactions is $\epsilon \leq
0.8\%$. \citet{Schneider06} showed that this value is too low to
explain such high energies of $E \geq 3\times 10^{61}$ erg.
Accretion is the only remaining mechanism which can yield larger
$\epsilon$. Its maximum efficiency is about $\epsilon \sim 6\%$ for
a non-rotating black hole and $\epsilon \sim 29\%$ for a maximally
rotating one \citep{Schneider06}. Therefore, AGN release their
enormous energy by matter accreting towards, and falling into, a
central supermassive black hole.

If the cold matter, which was initially at rest and without magnetic
field, was subjected to free radial infall it would accrete to the
central black hole without any energy release or observational
effects \citep{Shakura73}. However, in the case of AGN the accreting
matter has a significant angular momentum which does not allow its
free infall. At the marginally stable orbit of central black hole
with mass of $10^8M_\odot$, the specific angular momentum is
$\approx 1\times 10^{24}\ \mathrm{cm}^2/\mathrm{s}$, which is much
less even in comparison to a typical galaxy where the speciffic
angular momentum of orbiting material is $\approx 6\times 10^{28}\
\mathrm{cm}^2/\mathrm{s}$ \citep{Krolik99}. It means that approach
of accreting material toward a black hole requires a loss of the
greatest fraction of its initial angular momentum. Mechanisms which
can contribute to such loss of angular momentum are viscosity,
nonaxisymmetric gravitational forces, magnetic forces, etc.
\citep{Krolik99}.

Since the orbit of minimum energy for fixed angular momentum in any
spherically symmetric potential is a circle, the infall of accreting
material due to loss of its angular momentum will be in form of
successively smaller and smaller concentric circles. Matter
traveling along orbits inclined to each other will eventually
collide in the plane of intersection, and as a result, the angular
momenta of different gas steams will be mixed and equalized.
Consequently, all accreting matter will orbit in a single plane and
will have the same specific angular momentum at any given radius,
meaning that accretion is most likely performed through an accretion
disk \citep{Krolik99}.

The accreting material is able to approach the marginally stable
orbit of central black hole only if there is an efficient mechanism
for transporting angular momentum outward. Magnetic field that
exists in matter flowing into the disk, as well as its turbulent
motions, enable angular momentum to be transferred outward. Hence,
the accretion disk is one of the best candidates for a such
mechanism, since the particles in it lose their angular momentum due
to friction between its adjacent layers and spiral towards black
hole, releasing their gravitational energy. Part of this energy
increases the kinetic energy of rotation and the rest is turned into
the thermal energy and irradiated from the disk surface
\citep{Shakura73}.

The efficiency of angular momentum transport in the disk is
characterized by viscosity parameter $\alpha\le 1$, which can be
assumed as constant in most cases, since the spectrum of disk
radiation and its surface temperature do not strongly depend on it,
except in the case of a supercritical accretion regime
\citep{Shakura73}.

\subsection{Accretion rate and luminosity of AGN}

The total energy release and the spectrum of emitted radiation are
mainly determined by the rate of matter inflow into the disk on its
outer boundary, i.e. by its accretion rate which is usually denoted
by $\dot{M}$. If the accretion converts matter to radiation with
fixed radiative efficiency $\eta$ (in rest-mass units), then a
characteristic scale for the accretion rate is the Eddington
accretion rate \citep{Krolik99,Shakura73}:
\begin{equation}
\dot{M}_E=3\times 10^{-8}~\frac{0.06}{\eta }\frac{M}{M_\odot }\frac{M_\odot}{\mathrm{yr}},
\label{eq41_1}
\end{equation}
for witch the total release of energy in the disk $L=\eta\dot{M}c^2$
is equal to the Eddington luminosity:
\begin{equation}
L_E = 1.51\times 10^{38} \frac{M}{{M_ \odot  }}\mathrm{\frac{{erg}}{s}}.
\label{eq41_2}
\end{equation}
Eddington luminosity defines a critical luminosity for any given
mass, beyond which the radiation force overpowers gravity.

Observed AGN have luminosities from $\sim 10^{42}$ to $\sim 10^{48}$
erg/s, which means that their central black holes must have masses
from $10^5$ to $10^9\ M_\odot$, respectively \citep{Krolik99}. If we
assume that accretion in AGN occurs with rate of $\dot{M}\sim 0.5\
M_\odot/\mathrm{yr}$ and with radiative efficiency $\eta\sim 0.1$,
the resulting luminosity would be $\sim 3\times 10^{45}$ erg/s,
which is the middle of the observed luminosity distribution of low
redshifted AGN \citep{Krolik99}.

Actual accretion rates $\dot{M}$ of matter inflow into the accretion
disks of AGN could be many times less or higher than the critical
value of the Eddington accretion rate. At essentially subcritical
accretion rates ($\dot{M} \ll \dot{M}_E$), the maximal surface
temperatures $T_s$ are on the order of $10^5-10^6$ K in the inner
regions of the disk, from which the most of energy is released,
mainly in UV and soft X-ray bands \citep{Shakura73}. When the
accretion rate increases, the luminosity also raises, as well as the
effective radiation temperature. At accretion rates comparable to
$\dot{M}_E$, the accretion disk becomes a powerful source of X-ray
radiation with the effective radiation temperature $T_r$ from $10^7$
to $10^8$ K. In a strongly supercritical regime of accretion
($\dot{M} \gg \dot{M}_E$), the luminosity becomes fixed at the
Eddington critical limit $L_E$ and the most of energy is irradiated
from accretion disk in UV and optical spectral bands
\citep{Shakura73}.

\subsection{Standard model and spectral distribution}

By the term "standard Newtonian model of accretion disk", here we
will assume the disk model given by \citet{Shakura73}. This model
was originally developed to describe the accretion disks around
stellar sized black holes in the binary systems, but with certain
modifications it could be also applied on accretion disks around
supermassive black holes in AGN.

The standard Newtonian model  of accretion disk is based on the
supposition that the released gravitational energy is emitted as a
multitemperature blackbody radiation, where the surface temperature
profile is given by \citep[see][]{Shakura73}:
\begin{equation}
T_s(r) = \left[\frac{3}{8\pi} \frac{GM}{\sigma r^3} \dot{M}
\left(1 - \sqrt{\frac{r_{in}}{r}}\right)\right]^{1/4} ,
\label{eq42_1}
\end{equation}
where $\sigma$ is the Stefan constant, $G$ is the gravitation
constant, $M$ is the mass of the central black hole, $\dot{M}$ is
the accretion rate and $r_{in}$ is the inner radius of the accretion
disk. In above Equation and throughout this section we will use the
dimensionless disk radius $r$ defined as \citep{Popovic06a}:
\begin{eqnarray}
r=\frac{R}{6\ R_g}=
\frac{1}{6}\frac{R c^2}{GM}=
\frac{M_\odot}{M}\frac{R}{9~{\rm km}},
\label{eq42_0}
\end{eqnarray}
where $R$ is disk radius expressed in gravitational radii $R_g$.

In the standard model of accretion disk, accretion occurs via an
optically thick and geometrically thin disk. The effective optical
depth in the disk is very high and photons are close to thermal
equilibrium with electrons \citep{Popovic06a}. The spectrum of
thermal radiation emitted from accretion disk surface depends on its
structure and temperature (and therefore on the distance to the
black hole), and can have several distributions such as
\citep{Shakura73}:
\begin{enumerate}
\item Planck distribution in the outer regions of the disk,
    where free-free and free-bound processes, as well as
    absorption in the lines of heavy elements broadened by the
    gas pressure, give the main contribution to the shape of
    emitted spectrum:
\begin{equation}
F\left( x \right) = \frac{{2\pi h}}{{c^2 }}\left( {\frac{{kT}}{h}} \right)^3 \frac{{x^3 }}{{e^x  - 1}},
\label{eq42_2}
\end{equation}
where $x=\dfrac{h\nu}{k T}$, $h$ is Planck and $k$ is Boltzman
constant,
\item more complex spectrum which passed through the homogeneous
    or exponentially varying medium in intermediate regions of
    the disk where Thomson scattering dominates and
\item Wien distribution in the inner regions where the
    comptonization processes strongly affect the shape of
    emitted spectrum:
\begin{equation}
F\left( x \right) \sim x^3 e^{-x}.
\label{eq42_3}
\end{equation}
\end{enumerate}

\begin{figure}[ht!]
\centering
\includegraphics[width=\textwidth]{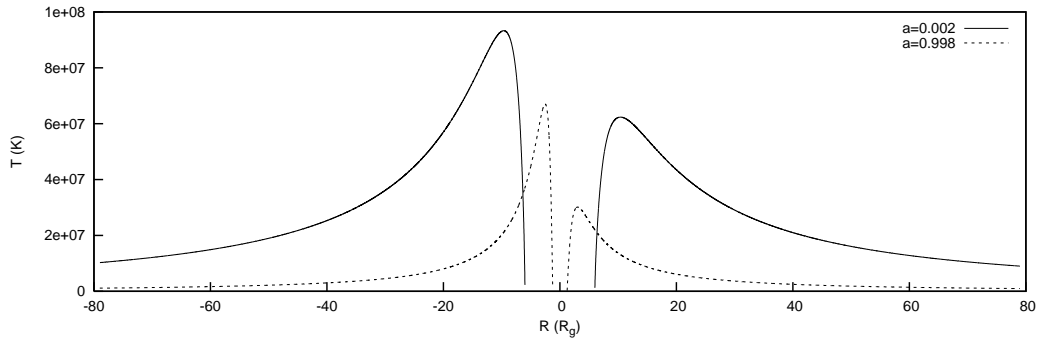}
\caption{The distribution of the temperature as a function of the
disk radius $R$ (in the direction normal to the rotation axis, as seen by distant observer), given for two
different values of angular momentum  $a$ of central black hole \citep{Popovic06a}.
Negative values of $R$ correspond to the approaching and positive to the receding side of the disk.}
\label{fig42_1}
\end{figure}

The surface temperature has different radial distributions in
different parts of accretion disk and results in the multicolor
black body spectrum. In the inner regions of accretion disks of AGN
where their X-ray radiation is generated, the radial distribution of
surface temperature has the following form \citep{Popovic06b}:
\begin{equation}
T_s(r) = T_0\ r^{-3/2}(1-r^{-1/2})^{4/5} \,\mathrm{K},
\label{eq42_4}
\end{equation}
where the temperature constant $T_0$ is chosen so that the
corresponding effective radiation temperature is $T_{r}=10^7-10^8$
K. This modification of standard disk model is made, because the
surface temperature given by Eq. (\ref{eq42_1}) could not be
successfully applied in the case of accretion disks of AGN. The
distribution of the temperature as a function of the accretion disk
radius, according to Eq. (\ref{eq42_4}), is presented in Fig.
\ref{fig42_1} for two different values of angular momentum $a$ of
central black hole \citep{Popovic06a}.

One can see from Fig. \ref{fig42_1} that the temperature depends not
only on disk radius, but also on its other parameters, such as e.g.
angular momentum $a$. Emitters located at different radii in
accretion disk have different temperatures and therefore make
different contributions to the observed flux. Also, it is noticeable
that the temperatures at smaller radii of approaching side of the
disk are significantly higher than those on its receding side.

Intensity of radiation emitted from the total disk surface can be
obtained by integration of all local spectra emitted from the above
regions:
\begin{equation}
I_\nu = 4\pi \int\limits_{r_{in} }^{r_{out} } {F_\nu  } \left[ {T_S \left( r \right)} \right]rdr,
\label{eq42_5}
\end{equation}
where $r_{in}$ and $r_{out}$ are inner and outer radii of accretion
disk, respectively.

\begin{figure}[ht!]
\centering
\includegraphics[width=\textwidth]{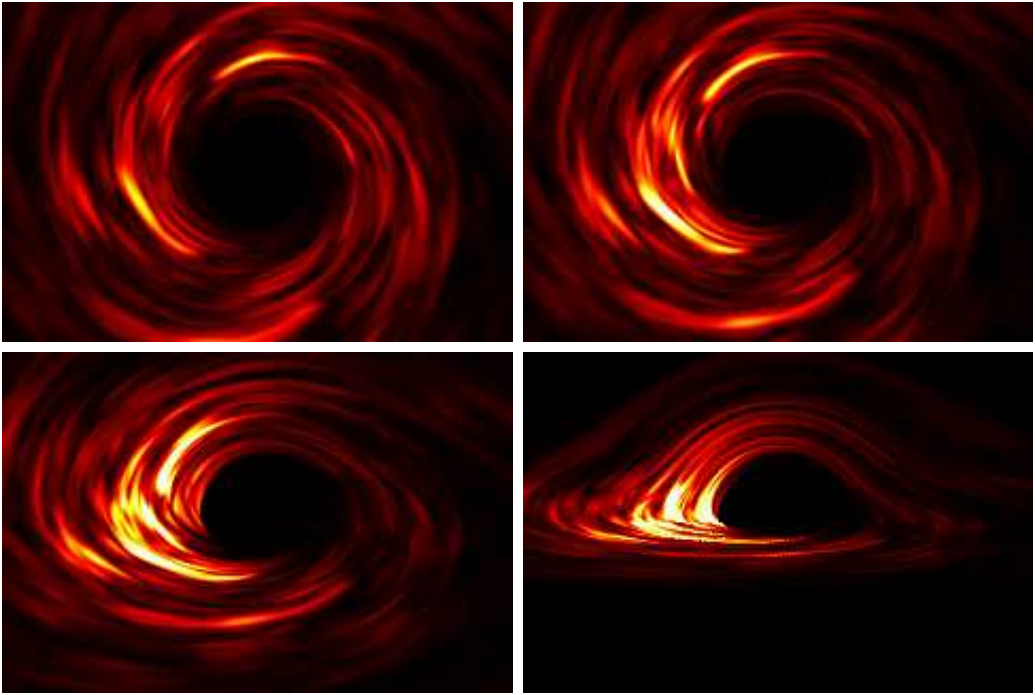}
\caption{Magnetohydrodynamic simulations of an accretion disk as seen by a distant observer at
the following inclination angles \citep{Armitage03}: 5$^\circ$ (upper left),
30$^\circ$ (upper right), 55$^\circ$ (lower left)
and 80$^\circ$ (lower right). The disk rotation is in counterclockwise direction.}
\label{fig42_2}
\end{figure}

Magnetohydrodynamic simulations of an accretion disk as seen by a
distant observer at four different inclinations are presented in
Fig. \ref{fig42_2} \citep{Armitage03}. The most distinctive details
in this figure are especially bright regions in form of arcs within
the turbulent flow, which trace out the photon trajectories close to
the radius of marginally stabile orbit. These arcs are much brighter
on approaching side of the disk due to higher surface temperature,
as shown in Fig. \ref{fig42_1}, and also due to Doppler boosting and
relativistic beaming which both enhance the flux observed in the
direction of motion and diminish the flux in the opposite
direction.\footnote{More details about the last two relativistic
effects can be found in e.g. \citet{Krolik99}}

\subsection{Structure and emission}

An accretion disk around a supermassive black holes in the center of
AGN extends from the radius of marginally stable orbit $R_{ms}$ to
the several thousands of gravitational radii. According to its
radiation emitted in different spectral bands, it can be stratified
in several parts \citep{Jovanovic08}:
\begin{enumerate}
\item innermost part close to the central black hole which emits
    X-rays and which extends from the radius of marginally
    stable orbit $R_{ms}$ to the several tens of gravitational
    radii \citep[see e.g.][]{Ballantyne05},
\item central part ranging from $\sim 100\ R_g$ to  $\sim 1000\
    R_g$ which emits UV radiation and
\item outer part extending from several hundreds to several
    thousands $R_g$, from which the optical emission is coming
    \citep{Eracleous94,Eracleous03}.
\end{enumerate}

Emissivity of the disk has an important role in the line and
continuum shapes. In the rest frame of the emitting material (i.e.
in disk frame), emissivity $\varepsilon(r)$ is defined as energy
emitted per unit proper time per unit proper area
\citep{Dabrowski97}. In the same frame, it is related to the emitted
intensity $I(r,\nu_e)$ and emitted flux $F(r,\nu_e)$ by
\citep{Dabrowski97}:
\begin{equation}
F(r,\nu_e) = \pi I(r,\nu_e) = \varepsilon(r) \delta(\nu_e-\nu_0),
\label{eq43_0}
\end{equation}
under assumption that the line emitted at frequency $\nu_0$ can be
approximated by a $\delta$ function.

Although the standard model of the accretion disk does not predict
the power-law for the disk emissivity, such law is usually accepted
in the case of the hard X-ray emission from the inner parts of AGN
accretion disks \citep{Fabian89,Nandra97,Nandra99}. In that case the
surface emissivity is assumed to vary with radius as power law with
emissivity index $p$ \citep{Fabian89,Fanton97}:
\begin{equation}
\varepsilon (r) = \dfrac{\varepsilon_0}{4\pi} r^{-p},
\label{eq43_1}
\end{equation}
where $\varepsilon_0$ is the emissivity constant. \citet{Fabian89}
found that emissivity index $p$ varies between 0 and 3 for the Fe
K$\alpha$ line emitting region of Cygnus X-1, where the most
probable value is $p\approx 2$. Some authors found that models
describing the disk emissivity as a broken power law, rather than a
single power law, achieve significantly better statistical fits of
the X-ray emission in the Fe K$\alpha$ line \citep{Brenneman06}. One
such broken power law between some inner radius $r_{in}$ and outer
radius $r_{out}$ with break radius $r_{br}$ is given by
\citep{Brenneman06}:
\begin{eqnarray}
\varepsilon(r)=\left\{
\begin{array}{ll}
0, & r<r_{in}\\
(r/r_{br})^{-\alpha_1}, & r_{in}\le r < r_{br} \\
(r/r_{br})^{-\alpha_2}, & r_{br}\le r < r_{out} \\
0, & r\ge r_{out}
\end{array}
\right. ,
\label{eq43_2}
\end{eqnarray}
where $\alpha_1$ and $\alpha_2$ are some emissivity indices. Using
this law \citet{Brenneman06} fitted the Fe K$\alpha$ line of
MCG--6-30-15, assuming a maximally spinning ($a=0.998$) black hole,
and found that the line emissivity followed such dependence for
$\alpha_1=4.5-6$ and $\alpha_2 \sim 2.5$, where the break radius was
$r_{br} \approx 6\ R_g$.

The X-ray continuum can be also modeled e.g. by the following
time-independent intrinsic emissivity of the disk as a function of
its radius $r$ and photon energy $E$ \citep{Dovciak04,Popovic06a}:
\begin{equation}
I(E,r)\sim E^{-\Gamma}\times r^{-\alpha},
\label{eq43_2b}
\end{equation}
where $\Gamma$ is photon index and $\alpha$ is radial emissivity
power law index of a continuum emitting region. According to some
studies of observed X-ray spectra of AGN, the values of $\Gamma$ and
$\alpha$ are estimated to be about 1.5 and 2.5, respectively
\citep{Dovciak04,Popovic06a}.

In the case of the disk outer parts (i.e. for optical emitting
region), the black-body emissivity law is assumed and, according to
the Eq. (\ref{eq42_2}), emitted intensity is given by Planck
function \citep{Popovic06a}:
\begin{equation}
I(E,T_s) = {\frac{{2E^{3}}}{{h^2c^{2}}}}{\frac{{1}}{{e^{{{E}
\mathord{\left/ {\vphantom {{h\nu}  {kT}}} \right.\kern-\nulldelimiterspace} {kT_s}}} - 1}}}.
\label{eq43_3}
\end{equation}

However, in the inner parts of the accretion disk Planck function
cannot be used properly. In these regions so called "modified"
black-body emissivity law is applied. According to the Eq.
(\ref{eq42_3}), the corresponding emitted intensity is given by
\citep{Popovic06a}:
\begin{equation}
I(E,T_s) \propto x^3 e^{-x},\hspace*{1cm} x=\frac{E}{kT_s}.
\label{eq43_4}
\end{equation}

\begin{figure}[ht!]
\centering
\includegraphics[width=0.8\textwidth]{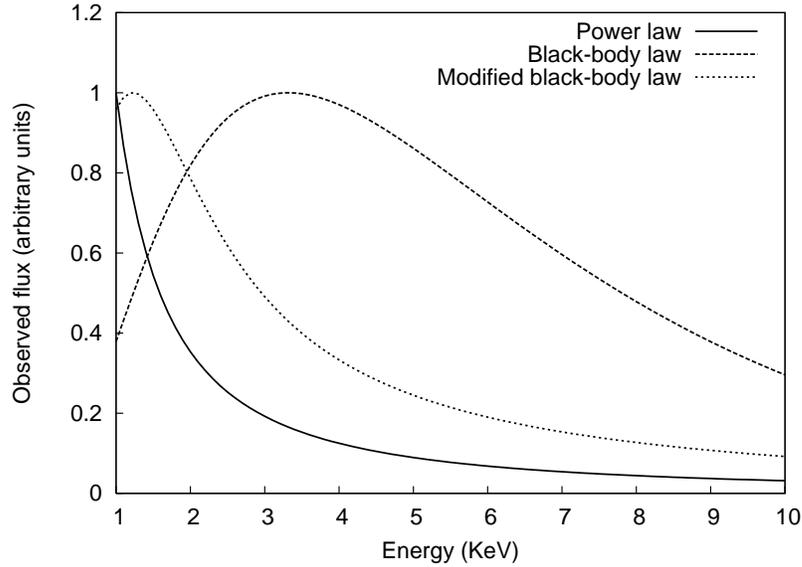}
\caption{Shapes of the X-ray continuum in 1 -- 10 keV energy band, emitted from an accretion disk
with the inner and outer radii equal to $R_{in} = R_{ms}$ and $R_{out}=80\ R_g$, around a supermassive black hole with mass $M=10^8\ M_\odot$
\citep{Popovic06a}. The continuum shapes are obtained assuming the following three emissivity laws (all normalized to their maximum values):
the power law defined by Eq. (\ref{eq43_1}) with emissivity index $p=2.5$, the black-body and "modified"
black-body emissivity laws defined by eqs. (\ref{eq43_3}) and (\ref{eq43_4}), respectively, assuming the radial distribution of the surface temperature given by Eq. (\ref{eq42_4}).}
\label{fig43_1}
\end{figure}

A comparison between the shapes of the X-ray continuum (1 -- 10 keV)
emitted from an accretion disk around a supermassive black hole,
obtained assuming three different emissivity laws, is presented in
Fig. \ref{fig43_1} \citep{Popovic06a}. From Fig. \ref{fig43_1} it is
clear that X-ray continuum shape strongly depends on adopted
emissivity law.

\section{Supermassive Black Holes and X-ray Emis\-sion}
\markright{Supermassive Black Holes and X-ray Emis\-sion}

A disk geometry for the X-ray emitting regions of AGN could be
assumed, since the unified model of AGN includes a supermassive
black hole fed by an accretion disk. \citet{Fabian89} calculated
spectral line profiles for radiation emitted from inner parts of
accretion disks and later on such features of Fe $K\alpha$ lines
were discovered by \citet{Tanaka95} in Japanese \emph{ASCA}
satellite data for Seyfert galaxy MGC-6-30-15. The assumption of a
disk geometry for the distribution of the X-ray emitters in the
central parts of AGN is also supported by the spectral shape of the
Fe K$\alpha$ line \citep{Nandra97,Nandra99,Nandra07}. Moreover, a
bump in the UV spectra of AGN is present very often, indicating that
the UV and optical continua also originate from an accretion disk
\citep{Jovanovic08}.

We should note here that probably most of the X-ray emission in the
1--10 keV energy range originates from inverse Compton scattering of
photons from the disk by electrons in a tenuous hot corona
\citep{Jovanovic08}. Proposed geometries of the hot corona of AGN
include a spherical corona sandwiching the disk and a patchy corona
made of a few compact regions covering a small fraction of the disk
\citep{Malzac07}. On the other hand, it is known that part of the
accretion disk that emits in the 1--10\ keV rest-frame band (e.g.
the region that emits the continuum Compton reflection component and
the fluorescent emission lines) is very compact and may contribute
to the X-ray variability in this energy range.

\begin{figure}[ht!]
\centering
\includegraphics[height=0.8\textwidth,angle=270]{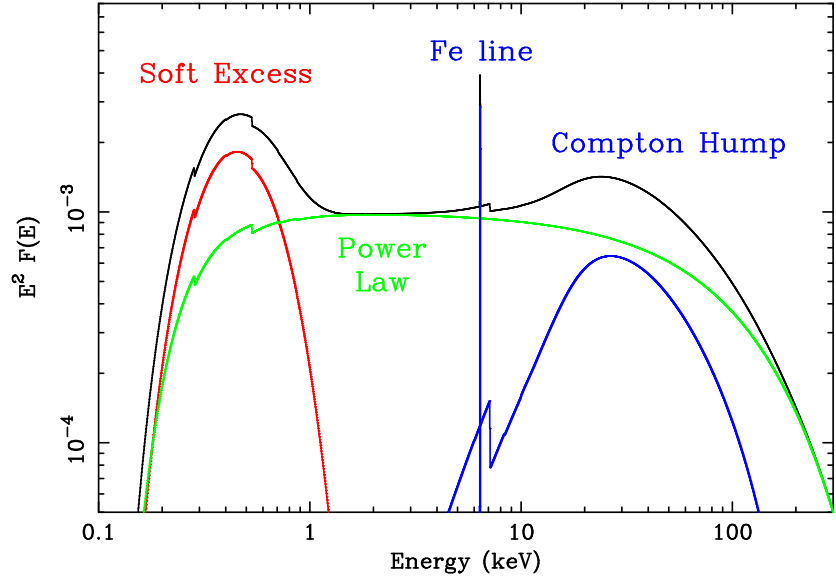}
\caption{A typical X-ray spectrum of an AGN \citep{Fabian06}.}
\label{fig50_1}
\end{figure}

A typical X-ray spectrum of AGN  (see Fig. \ref{fig50_1}) is
composed from the following components \citep{Fabian06}:
\begin{list}{-}{}
\item an underlying power-law continuum due to thermally Comptonized
    soft photons,
\item a soft excess at low energies below 1 keV due to thermal
    (black-body) emission from an optically-thick accretion disk,
\item a fluorescent/recombination Fe K$\alpha$ line and
\item a Compton hump due to X-ray reflection from the disk.
\end{list}

In the following text we will discuss in more details power law
component of X-ray continuum and the Fe $K\alpha$ spectral line.

\subsection{X-ray continuum of AGN}

\begin{figure}[ht!]
\centering
\includegraphics[height=0.8\textwidth,angle=270]{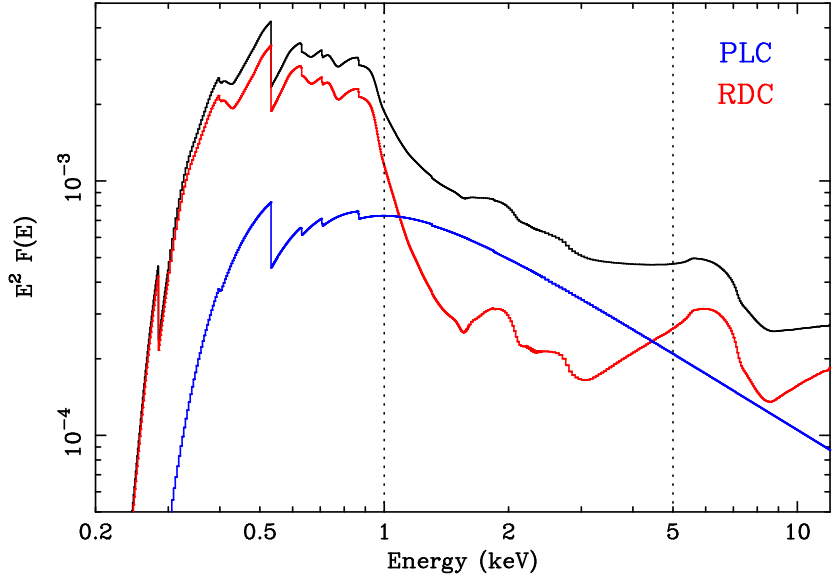}
\caption{The two-component model of the X-ray continuum in the case of Seyfert 1 galaxy 1H 0707--495, composed from
a power law  (PLC) and a reflection (RDC) component \citep{Fabian06}.}
\label{fig51_1}
\end{figure}

AGN are powerful sources of X-ray radiation in the continuum from
0.1 to 100 keV, which contains two components (see Fig.
\ref{fig51_1}): the soft X-rays with a steep spectrum and the hard
component with spectrum in form of a power law
\citep{Fabian89,Fabian06}. Therefore, the observed continuum flux is
very often fitted with one or two black-body components in the soft
X-rays, in addition to a power law component in the hard X-rays. It
is believed that both, soft and hard components mostly arise from
the inner region of AGN disk, close to the central supermassive
black hole ($\sim 10\ R_S$). The first component is probably formed
in the accretion disk, which is also a strong source of the soft UV
and optical photons, or at least in cold ($T<10^7$ K) accreting gas
clouds. The second one is caused by high-energy (most likely
relativistic) electrons in hot corona above the disk, when they
multiply inverse-Compton scatter some of the low-energy UV and
optical photons from the disk to X-ray energies
\citep{Fabian89,Fabian00}. The resulting hard X-ray power law
component irradiates the accretion disk and produces a reflection
component which causes the observed spectrum to flatten above 10 keV
(see Fig. \ref{fig51_1}), as Compton recoil reduces the
backscattered flux \citep{Fabian00}. Apart from its own thermal
radiation, the cold gas around a black hole is also irradiated by
these hard X-ray photons, producing different spectral features
through photoelectric absorption, fluorescence (responsible for
occurrence of fluorescent iron lines) and scattering
\citep{Fabian89}. Therefore, the observed X-ray spectra of AGN
consist of an iron line spectrum superposed on a power law continuum
\citep{Fabian00}.

Fluctuations of the X-ray radiation on timescales from several parts
of an hour until several days are a common property of all AGN
\citep{Krolik99}. Such fast variations confirm assumption that X-ray
radiation is emitted from a very compact region in the center of
AGN.

\citet{Fabian06} showed that a two-component model, consisting of a
power-law and a reflection component, could better describe the
observed variability of X-ray spectra in the case of several bright
Seyferts than a simple power-law model. This two-component model is
presented in Fig. \ref{fig51_1}.

\subsection{Fe K$\alpha$ spectral line}

\begin{figure}[ht!]
\centering
\includegraphics[width=\textwidth]{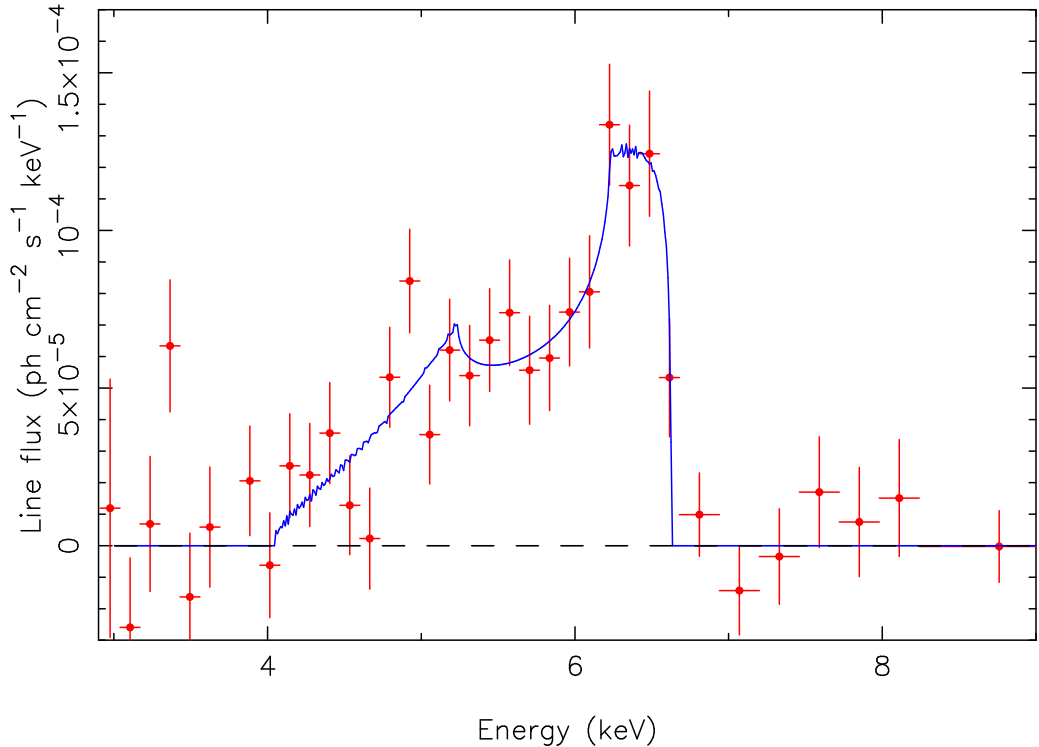}
\caption{The profile of the Fe K$\alpha$ line from Seyfert 1 galaxy MCG-6-30-15 observed by
\emph{ASCA SIS} detector \citep{Tanaka95} and the best fit (blue solid line) obtained by
a model of the accretion disk in Schwarzschild metric, extending between 3 and 10 Schwarzschild radii \citep{Fabian89}.
Image Credit: Tanaka et al., Copyright: Nature, 375, 659, 1995.
Image generated by Dr. Paul Nandra NASA/GSFC.}
\label{fig52_1}
\end{figure}

Contrary to the UV and optical spectra of AGN, there are very few
strong spectral lines in their X-ray spectra. The most important one
is the Fe K$\alpha$ line. Iron abundance in accretion disks of AGN
is sufficient to produce very strong emission in this line. The CCD
detectors on Japanese \emph{ASCA} satellite were the first
instruments with sufficient spectral resolution and sensitivity in
the X-ray band, by which \citet{Tanaka95} obtained the first
convincing proof for the existence of the Fe K$\alpha$ line in AGN
spectra. This discovery was made after four-day observations of
Seyfert 1 galaxy MCG-6-30-15 (see Fig. \ref{fig52_1}).

The fluorescent/recombination iron K$\alpha$ line is an important
indicator of accreting flows around compact objects, because it is
produced in inner parts of their accretion disks. At the same time,
it is the strongest line of the X-ray radiation, and it can be found
in the spectra of all types of accreting sources: binary black hole
and neutron star systems, cataclysmic variable stars and AGN.

The Fe K$\alpha$ line is produced when plasma is subjected to the
influence of the hard X-ray radiation so that one of the two
$K$-shell ($n=1$, where $n$ is the principal quantum number)
electrons of an iron atom (or ion) is ejected following the
photoelectric absorption of an X-ray \citep{Fabian00}. The threshold
for the absorption by neutral iron is $7.1$ keV \citep{Fabian00}.
The resulting excited state decays when an $L$-shell ($n=2$)
electron drops into the $K$-shell, releasing $6.4$ keV of energy.
This energy is either emitted as an emission-line photon (34\%
probability) or internally absorbed by another electron (66\%
probability) which is consequently ejected from the iron ion (Auger
effect).

The fluorescent yield (i.e. the probability that photoelectric
absorption is followed by fluorescent line emission rather than the
Auger effect) is a weak function of the ionization state from
neutral iron (Fe I) up to Fe XXIII \citep{Fabian00}. For
lithium-like iron (Fe XXIV) through to hydrogen-like iron (Fe XXVI),
the lack of at least two electrons in the $L$-shell means that the
Auger effect cannot occur. For He and H-like iron ions, the line is
produced by the capture of free electrons (recombination) and the
equivalent fluorescent yield is high and it depends on the plasma
conditions \citep{Fabian00}.

For the neutral iron, the Fe K$\alpha$ line energy is 6.4 keV (more
precisely, there are two components of the line \citep{Fabian00}: Fe
K$\alpha_1$ at 6.404 and Fe K$\alpha_2$ at 6.391 keV), while in the
case of ionization, the energy of both the photoelectric threshold
and the Fe K$\alpha$ line are slightly increased. Even for such high
ionization states of He and H-like iron ions, the Fe K$\alpha$ line
energy is increased only to $6.7$ and $6.9$ keV, respectively
\citep{Krolik99}.

Fe K$\alpha$ line is pretty narrow in itself, but in case when it
originates from a relativistically rotating accretion disk of AGN it
becomes wider due to kinematical effects, and also its shape (or
profile) is changed due to Doppler boosting and gravitational
redshift. Such broadening of the line is very often observed in
spectra of Seyfert galaxies and is one of the main evidences for the
existence of a relativistic accretion disk which extends deeply in
the gravitational field of the central black hole \citep{Zycki04}.
If the line originated from an arbitrary radius of a nonrelativistic
(Keplerian) accretion disk it would have a symmetrical profile (due
to Doppler effect) with two peaks: a "blue" one which is produced by
emitting material from the approaching side of the disk in respect
to an observer, and a "red" one which corresponds to emitting
material from the receding side of the disk. The widest parts of the
Fe K$\alpha$ line arise from the innermost regions of the disk,
where the rotation of emitting material is the fastest. Using
\emph{ASCA} satellite observations, \citet{Nandra97} found that, in
case of 14 Seyfert 1 galaxies, Full-Widths at Half-Maximum (FWHM) of
their Fe K$\alpha$ lines correspond to velocities of $\approx
50,000$ km/s. In some cases (e.g. for Seyfert 1 galaxy MCG-6-30-15),
FWHM velocity reaches 30\% of speed of light \citep[see
e.g.][]{Nandra07}. It means that in the vicinity of the central
black hole, orbital velocities of the emitting material are
relativistic, causing the enhancement of the Fe K$\alpha$ line
"blue" peak in regard to its "red" peak (relativistic beaming).
Taking into account the integral emission in the line over all radii
of accretion disk, one can obtain the line with asymmetrical and
highly broadened profile \citep{Fabian00}. The "blue" peak is then
very narrow and bright, while the "red" one is wider and much
fainter (see Fig. \ref{fig52_1}). Besides, the gravitational
redshift causes further deformations of the Fe K$\alpha$ line
profile by smearing the "blue" emission into "red" one. Since the
observed Fe K$\alpha$ line profiles are strongly affected by such
relativistic effects, they represent a fundamental tool for
investigating the plasma conditions and the space-time geometry in
the vicinity of the supermassive black holes of AGN.

One of the important features of the Fe K$\alpha$ line is
variability of both, its shape and intensity. Observed variations of
this line are surprisingly less than those of the high energetic
continuum, which is assumed to give rise to the line emission
\citep{Zycki04}. Also, it seems that there is a lack of
corresponding line response to the continuum variations on time
scales from several minutes to several days, or that these line and
continuum variations are uncorrelated \citep{Zycki04}. In \S 6 of
this chapter we will pay attention to some possible causes of such
behavior of the Fe K$\alpha$ line and X-ray continuum.

\subsection{Modeling of X-ray emission using ray-tracing in Kerr metric}

\begin{figure}[ht!]
\centering
\includegraphics[width=\textwidth]{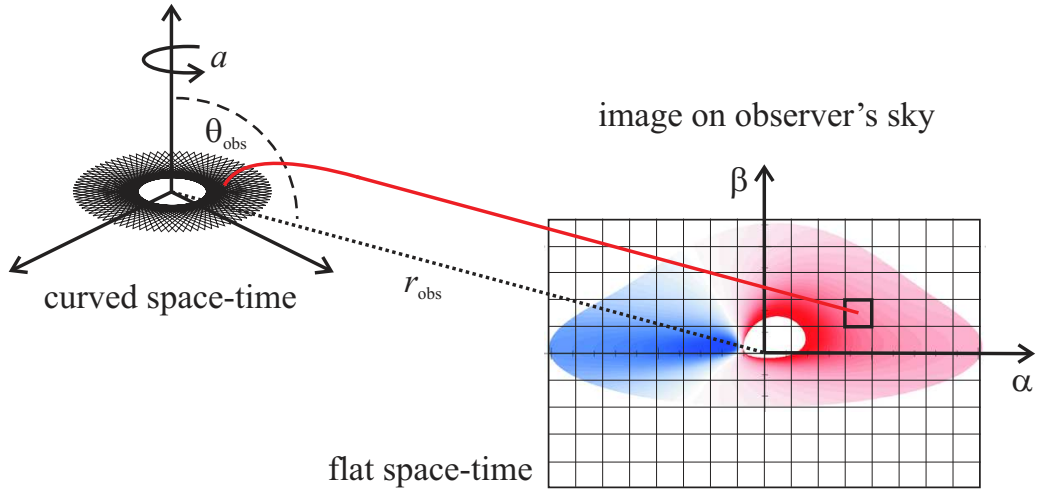}
\caption{Schematic illustration of the ray-tracing method in the Kerr metric, showing a light ray emitted
from some radius of accretion disk in coordinate system defined by a rotating black hole with angular momentum $a$,
and observed at a pixel with coordinates (impact parameters)
$\alpha,\beta$ on the disk image in the observer's reference frame (Figure courtesy: Vesna Borka Jovanovi\' c).}
\label{fig53_1}
\end{figure}

The disk emission can be analyzed by numerical simulations, based on
so called ray-tracing method in Kerr metric
\citep{Bao94,Bromley97,Fanton97,Cadez98}, taking into account only
photon trajectories reaching the observer's sky plane. In this
method one divides the image of the disk on the observer's sky into
a number of small elements (pixels). For each pixel, the photon
trajectory is traced backward from the observer by following the
geodesics in a Kerr space-time, until it crosses the plane of the
disk (see Fig. \ref{fig53_1}). Then, the flux density of the
radiation emitted by the disk at that point, as well as the redshift
factor of the photon are calculated. In that way, one can obtain the
color images of the accretion disk which a distant observer would
see by a high resolution telescope. The simulated line profiles can
be calculated taking into account the intensities and received
photon energies of all pixels of the corresponding disk image. All
illustrations of the accretion disk and the line shape in this
chapter are obtained using such numerical simulations. Here we will
briefly describe a pseudo-analytical approach of ray-tracing
proposed by \citet{Cadez98}.

This method is based on the pseudo-analytical integration of the
geodesic equations which describe the photon trajectories in the
general case of a rotating black hole having some angular momentum
$J$, which gravitational field is therefore described by the Kerr
metric (see \S 3.2 for more details). The Kerr metric depends on the
angular momentum normalized to the mass $M$ of black hole: $a=J/Mc,\
0\le a \le M$.

A photon trajectory in the Kerr metric can be described by three
constants of motion (the energy at infinity and two constants
related to the angular momentum, respectively) which have the
following forms when natural units $c=G=M=1$ are assumed
\citep{Cadez98}: $E=-p_t$, $\Lambda=p_\phi$ and $Q=p^2_\theta-a^2
E^2 cos^2\theta+\Lambda^2 cot^2\theta$. Here, $(r,\theta,\phi,t)$
are the usual Boyer-Lindquist coordinates and $p$ is the 4-momentum.
As the trajectory of a photon is independent on its energy, it may
be expressed using the two dimensionless parameters
$\lambda=\Lambda/E$ and $q=Q^{1/2}/E$ which are very simply related
to the two impact parameters $\alpha$ and $\beta$ describing the
apparent position on the observer's celestial sphere: $\alpha = -
{\dfrac{{\lambda}} {{\sin \theta _{obs}}} } $ and $ \beta = \pm
\left( {q^{2} + a^{2}\cos ^{2}\theta _{obs} - \lambda ^{2}\cot
^{2}\theta _{obs}} \right)^{{\frac{{1}}{{2}}}}$, where the sign of
$\beta$ is determined by $\left( {{\dfrac{{dr}}{{d\theta}} }}
\right)_{obs}$.

In order to find the photon trajectories (null geodesics) which
originate in the accretion disk at some emission radius $r_{em}$ and
reach the observer at infinity, one must solve the following
integral equation \citep{Cadez98}:

\begin{equation}
\pm \int\limits_{r_{em}} ^{\infty}  {{\dfrac{{dr}}{{\sqrt {R\left(
{r,\lambda ,q} \right)}}} }}  = \pm \int\limits_{\theta _{em}} ^{\theta
_{obs}}  {{\dfrac{{d\theta}} {{\sqrt {\Theta \left( {\theta ,\lambda ,q}
\right)}}} }},
\label{eq53_1}
\end{equation}

\begin{equation}
\begin{array}{c}
R\left( {r,\lambda ,q} \right) = \left( {r^{2} + a^{2} - a\lambda}
\right)^{2} - \Delta {\left[ {\left( {\lambda - a} \right)^{2} + q^{2}}
\right]}, \\
\Theta \left( {\theta ,\lambda ,q} \right) = q^{2} + a^{2}\cos ^{2}\theta -
\lambda ^{2}\cot ^{2}\theta .
\end{array}
\label{eq53_2}
\end{equation}

Above integral Equation (\ref{eq53_1}) can be solved in terms of
Jacobian elliptic functions, and therefore it is a pseudo-analytical
integration. For the exact expressions of the solutions, see e.g.
\citet{Cadez98}, or more recent paper from \citet{Li05}.

Due to relativistic effects, photons emitted at frequency $\nu_{em}$
will reach infinity at frequency $\nu_{obs}$, and their ratio
determines the shift due to these effects: $g = \dfrac{{\nu _{obs}}}
{{\nu _{em}}}$. The total observed flux at the observed energy
$E_{obs}$ is given by \citep{Fanton97}:
\begin{equation}
F_{obs} \left( {E_{obs}}  \right) = {\int\limits_{image} {\varepsilon \left(
{r} \right)}} g^{4}\delta \left( {E_{obs} - gE_{0}}  \right)d\Xi ,
\label{eq53_3}
\end{equation}
where $\varepsilon \left( {r} \right)$ is the disk emissivity,
$d\Xi$ is the solid angle subtended by the disk in the observer's
sky and $E_{0}$ is the rest energy.

A simulated accretion disk image is obtained in the following way:
\begin{enumerate}
\setlength{\itemsep}{-0.2cm}
\item values of the following input parameters are specified:
    inner and outer radii ($R_{in}$ and $R_{out}$) of the disk,
    angular momentum $a$ of the central black hole, observer's
    viewing angle (disk inclination) $\theta_{obs}$ (also, often
    denoted by $i$) and parameters defining the disk emissivity
\item constants of motion $\lambda$ and $q$ are calculated for
    each pixel on imaginary observer's photographic plate (i.e.
    for each pair of impact parameters $\alpha$ and $\beta$)
\item geodesic Equation (\ref{eq53_1}) is integrated for each
    pair of $\lambda$ and $q$
\item values of shift due to relativistic effects $g$ and
    observed flux $F_{obs}$ are calculated
\item pixels on imaginary observer's photographic plate are
    colored according to the value of shift $g$ and a simulated
    disk image is obtained.
\end{enumerate}

The simulated line profiles can be calculated from the corresponding
disk images by binning the observed flux at all pixels over the bins
of shift $g$. The examples of simulated disk images obtained in such
way are presented in left panels of Figs. \ref{fig54_1} -
\ref{fig54_3}, and the corresponding simulated line profiles are
presented in the right panels of the same figures.

\subsection{Observational effects of strong gravity in the vicinity of supermassive black holes}

In general, black holes have three measurable parameters (not
including the Hawking temperature): charge, mass (and hence
gravitational field) and angular momentum (or spin). In the case of
supermassive black holes of AGN, only the latter two are of
sufficient importance because they are responsible for several
effects which can be detected in the observed Fe K$\alpha$ line
shapes \citep{Jovanovic08a}.

\begin{figure}[ht!]
\centering
\includegraphics[width=0.48\textwidth,angle=270]{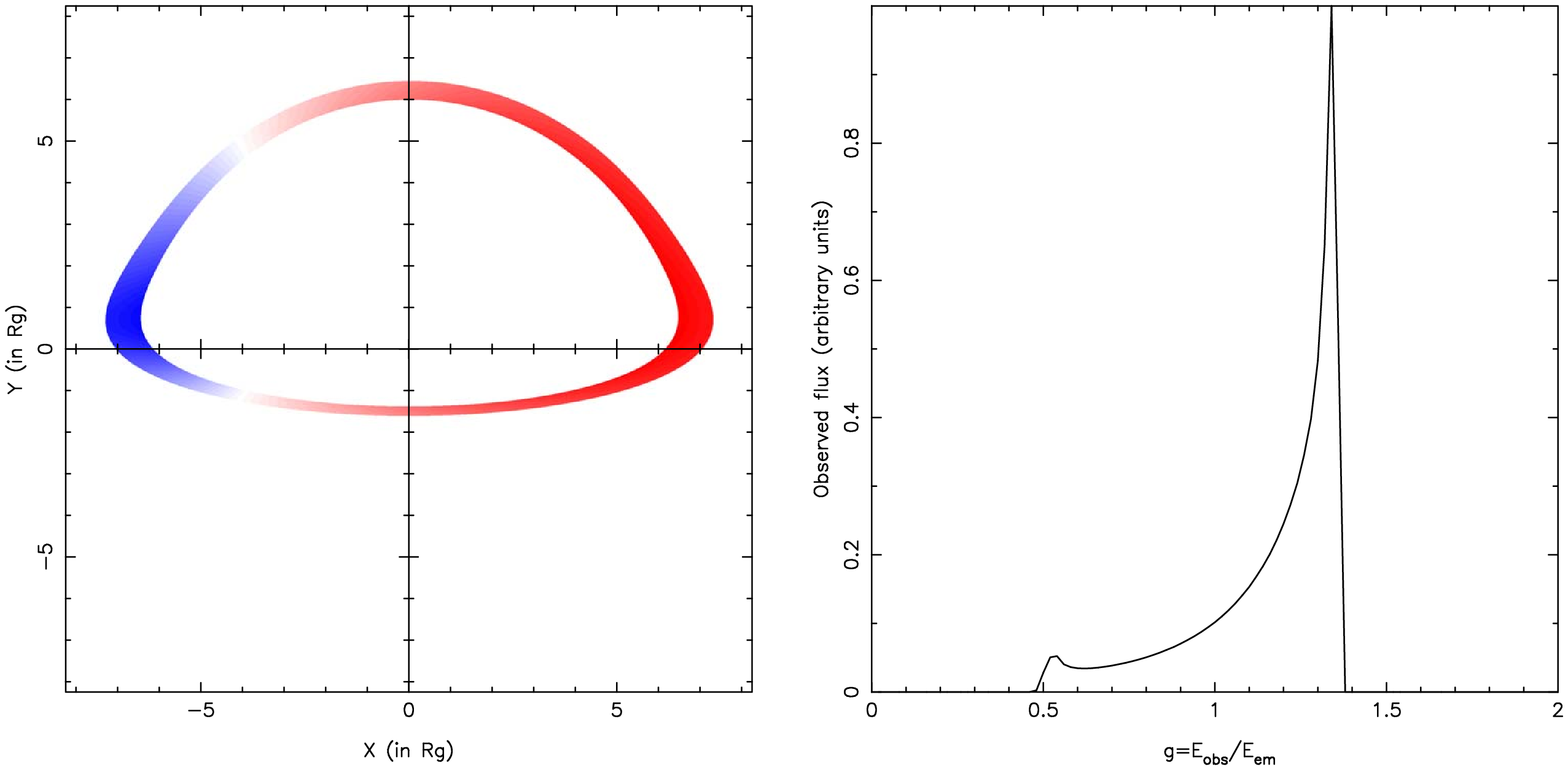} \\
\includegraphics[width=0.48\textwidth,angle=270]{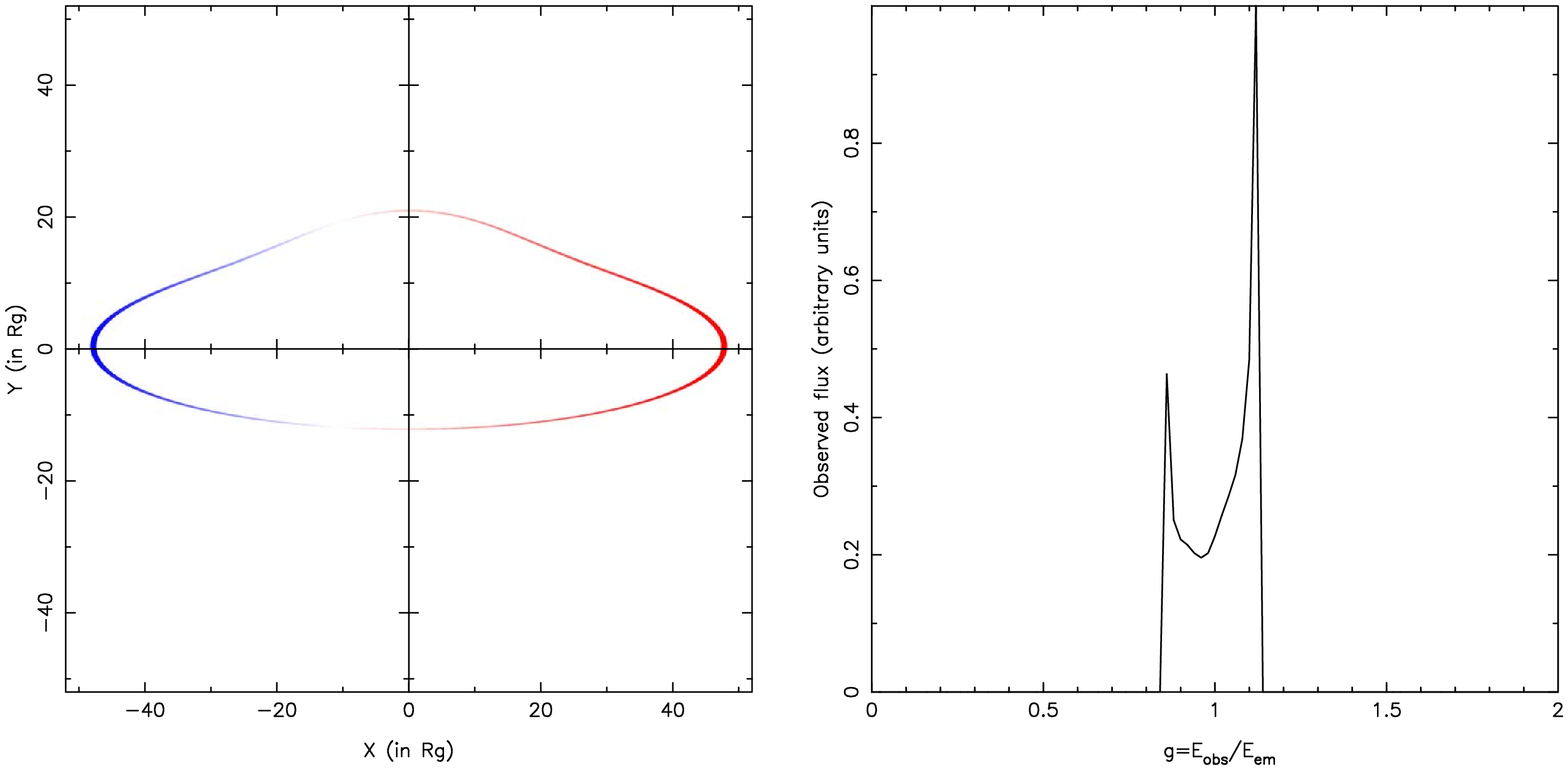}
\caption{\textit{Left:} illustrations of the Fe K$\alpha$ line emitting region in
form of narrow ring with width $=1 R_{g}$, extending from:
$R_{in}=6$ R$_{g}$ to $R_{out}=7$ R$_{g}$ (top) and $R_{in}=50$
R$_{g}$ to $R_{out}=51$ R$_{g}$ (bottom). \textit{Right:} the corresponding Fe K$\alpha$ line profiles \citep{Jovanovic08a}.}
\label{fig54_1}
\end{figure}

In order to study the size of the Fe K$\alpha$ line emitting region,
as well as its location in the disk, one can assume that the line is
emitted from a region in form of a narrow ring. For example,
\citet{Jovanovic08a} assumed a line emitting region with width
equals to $1\ R_{g}$, located between: a) $R_{in}=6$ R$_{g}$ and
$R_{out}=7$ R$_{g}$ and b) $R_{in}=50$ R$_{g}$ and $R_{out}=51$
R$_{g}$. These two cases are presented in Fig. \ref{fig54_1}.

From Fig. \ref{fig54_1} one can see how the Fe K$\alpha$ line
profile is changing as the function of distance from central black
hole. When the line emitters are located at the lower radii of the
disk, i.e. closer to the central black hole, they rotate faster and
the line is broader and more asymmetric (see Fig. \ref{fig54_1}
top-right). If the line emission is originating at larger distances
from the black hole, its emitting material is rotating slower and
therefore the line becomes narrower and more symmetric (see Fig.
\ref{fig54_1} bottom-right). In majority of AGN, where the broad Fe
K$\alpha$ line is observed, its profile is more similar to the
modeled profile as obtained under assumption that the line emitters
are located close to the central black hole
\citep{Tanaka95,Nandra07,Jovanovic08a}.

Angular momentum or spin of the central supermassive black hole of
AGN is a property of the space-time metric. To demonstrate how it
affects the observed line profiles we will now assume that the Fe
K$\alpha$ line emitting region extends between the following inner
and outer radii: $R_{in}=R_{ms}$ and $R_{out}=20\ R_g$. We will
analyze the following two cases for accretion disk inclination in
both Schwarzschild and Kerr metrics: (i) $i=35^\circ$ and (ii)
$i=75^\circ$. In the case of Schwarzschild metric we have a
stationary black hole and hence, angular momentum (normalized to the
mass of black hole) is $a=0$. In Kerr metric (i.e. for a rotating
black hole), it can take any value from the $[0, 1]$ range, but in
these two examples we will assume an almost maximally rotating black
hole with $a=0.998$.

\begin{figure}[ht!]
\centering
\includegraphics[width=\textwidth]{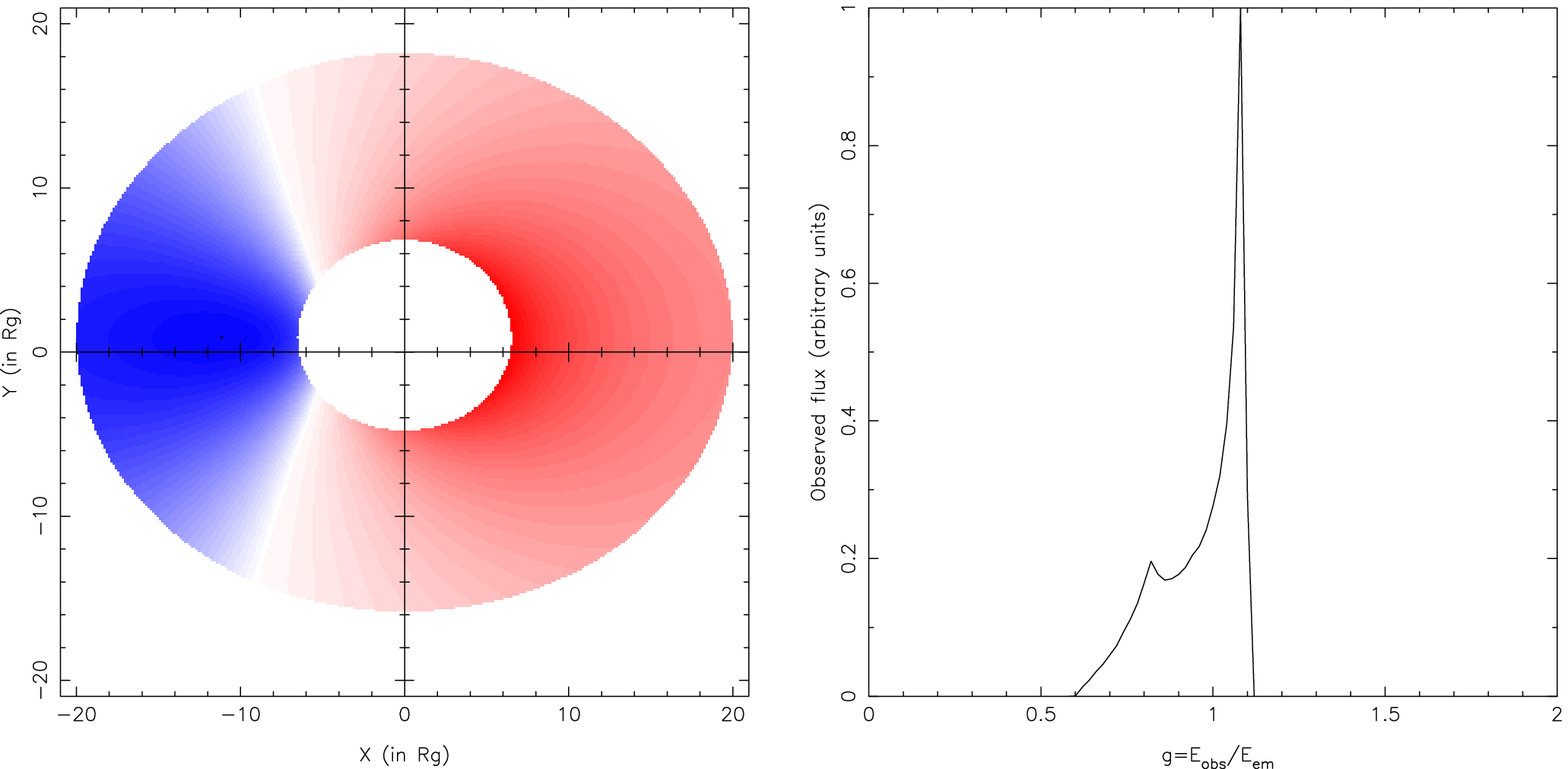} \\
\includegraphics[width=\textwidth]{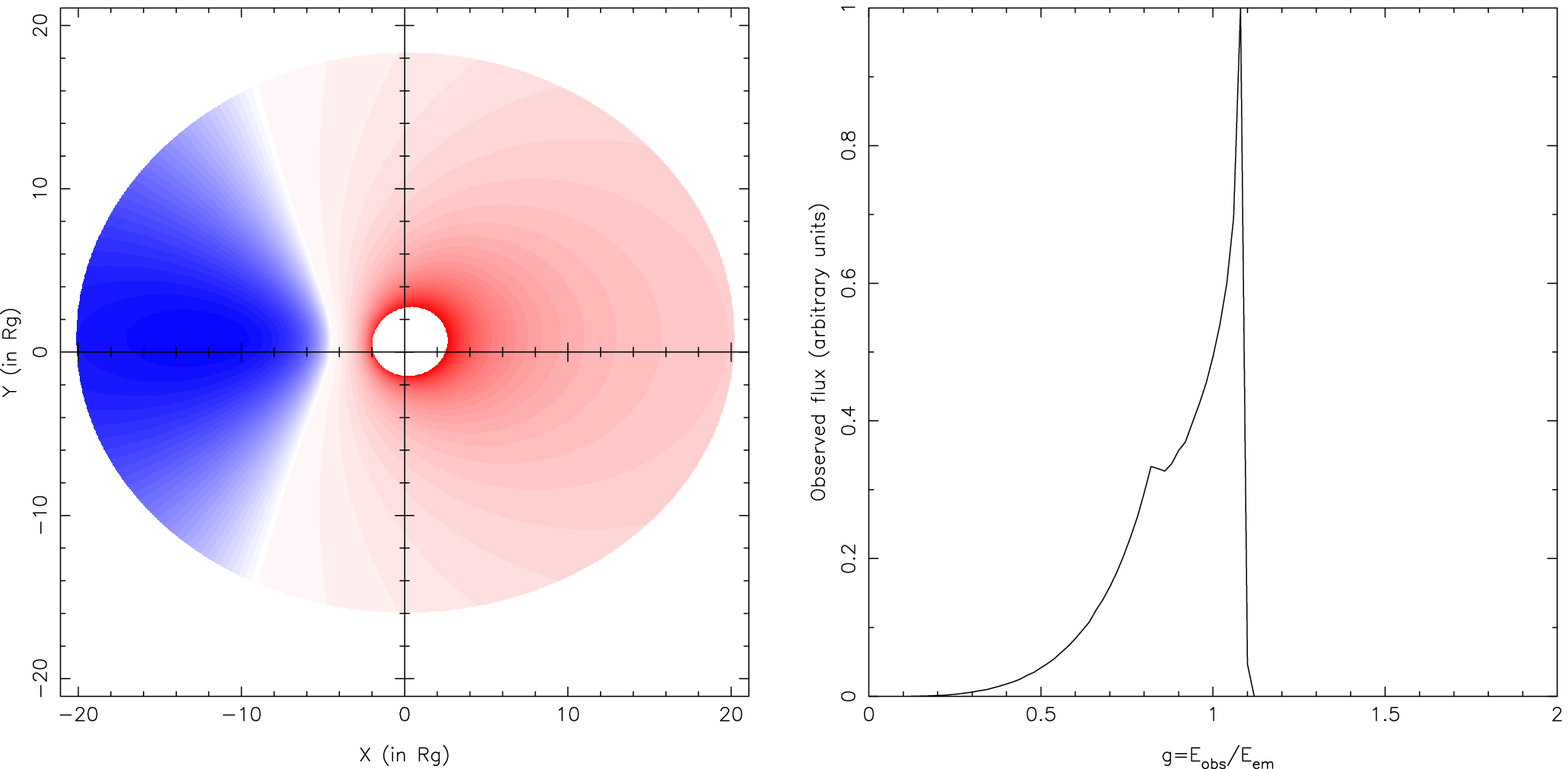}
\caption{Illustrations of accretion disk (left) and the
corresponding Fe K$\alpha$ line profiles (right) in the case of
Schwarzschild (top) and Kerr metric with angular momentum
$a=0.998$ (bottom). The disk inclination is $i=35^\circ$ and its
inner and outer radii are $R_{in}=R_{ms}$ and $R_{out}=20$ R$_{g}$,
respectively \citep{Jovanovic08a}.} \label{fig54_2}
\end{figure}

\begin{figure}[ht!]
\centering
\includegraphics[width=\textwidth]{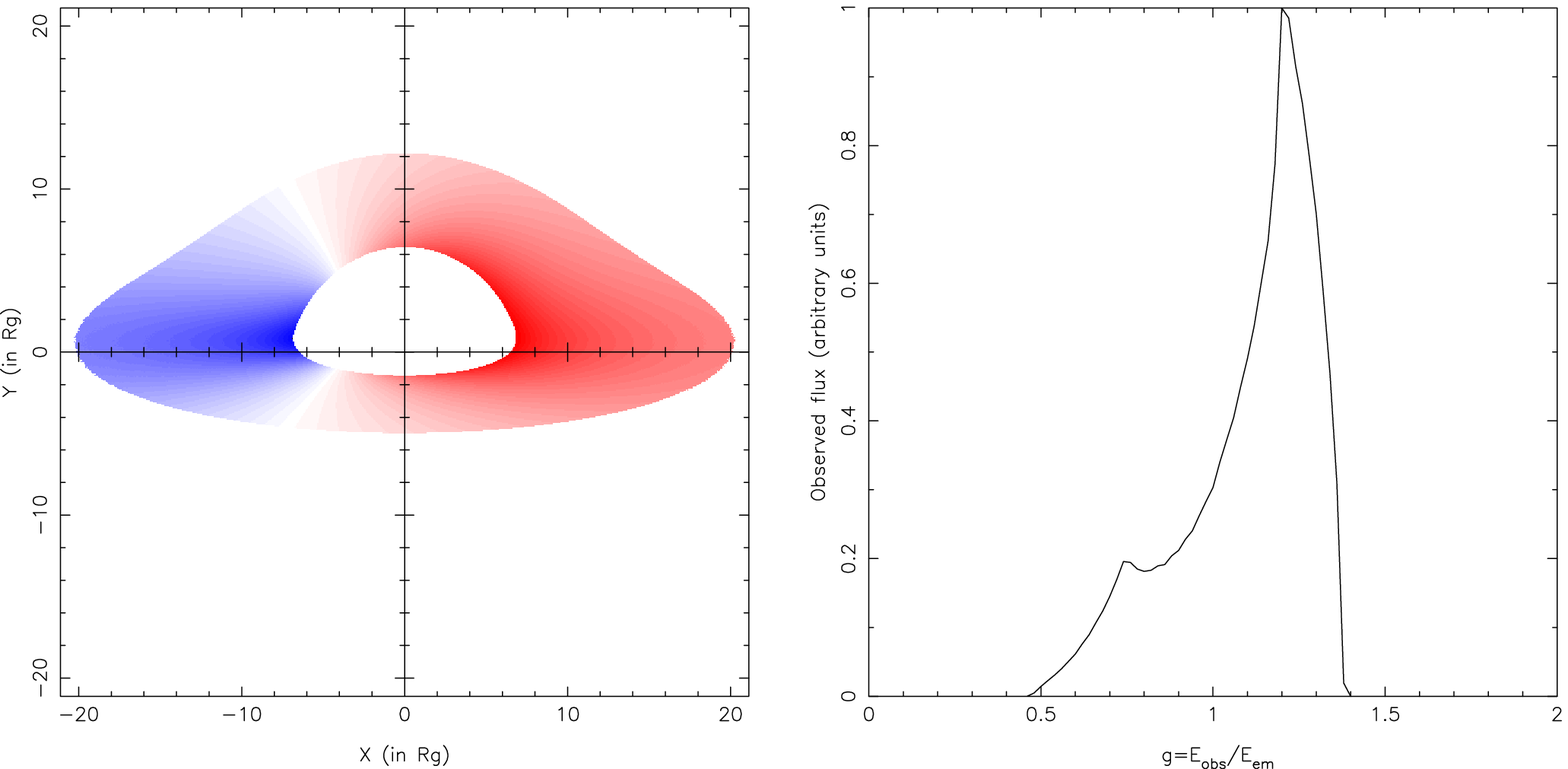} \\
\includegraphics[width=\textwidth]{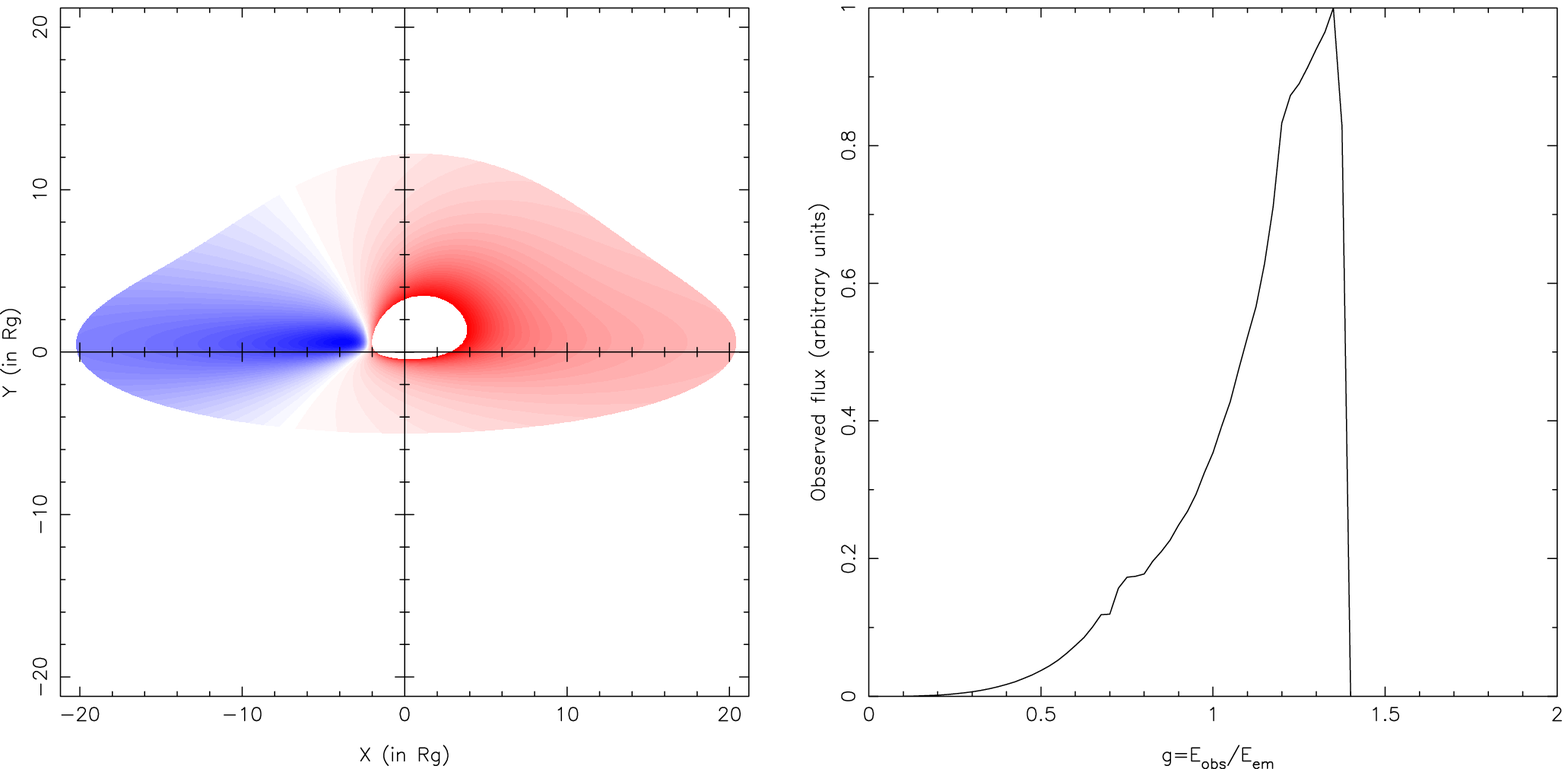}
\caption{The same as in Fig. \ref{fig54_2}, but for a highly inclined disk
with $i=75^\circ$ \citep{Jovanovic08a}.}
\label{fig54_3}
\end{figure}

Illustrations of an accretion disk and the corresponding Fe
K$\alpha$ line shapes in the first case are presented in Fig.
\ref{fig54_2}. As one can see from Fig. \ref{fig54_2}, the red peak
of the Fe K$\alpha$ line is brighter in case of almost maximally
rotating black hole, but at the same time it is also more embedded
into the blue peak wing and therefore less separable from it.

Fig. \ref{fig54_3} contains illustrations of the line emitting
regions and the corresponding line shapes in the case of a highly
inclined disk ($i=75^\circ$). Here, the line profiles are broader
than in the first case, mostly due to higher inclination. As it can
be seen from Fig. \ref{fig54_3}, in the case of the Kerr metric, the
red peak of the line is again more embedded into its blue peak wing
(as in the first case) and it confirms that this effect can be most
likely attributed to angular momentum \citep{Jovanovic08a}.
Consequently, angular momentum of the central black hole has
significant influence on the line shape which supports assumption
that the line originates from the innermost part of accretion disk,
close to the central black hole \citep[see e.g.][]{Ballantyne05}.
This fact can be used for estimation of angular momentum of the
central black hole in observed AGN.

Above simulations of the strong gravitational field influence on the
Fe K$\alpha$ line show that such effects can be detected in the
observed line shapes and therefore, comparisons between the observed
and modeled Fe K$\alpha$ line profiles (see e.g. Fig. \ref{fig52_1})
can bring us some essential information about the strong
gravitational field in the vicinity of central supermassive black
holes of AGN \citep{Nandra07}.

\clearpage

\section{Variability of X-ray Emission Around a Supermassive Black Ho\-le}
\markright{Variability of X-ray Emission Around a Supermassive Black
Hole}

Rapid and irregular variability of the observed X-ray emission in
the line, as well as in the continuum, is a common property of all
AGN. This variability could be due to disk instability, reflecting
in perturbations of the disk emissivity, or it could be caused by
some external effects, such as gravitational microlensing and
absorption by X-ray absorbers. We developed a model of perturbations
of the disk emissivity, a model of absorption region and three
models of gravitational microlensing. In the following text we will
pay more attention to all three mentioned causes of the X-ray
variability in AGN.

\subsection{Perturbations of disk emissivity}

\begin{figure}[ht!]
\centering
\includegraphics[width=0.8\textwidth,angle=270]{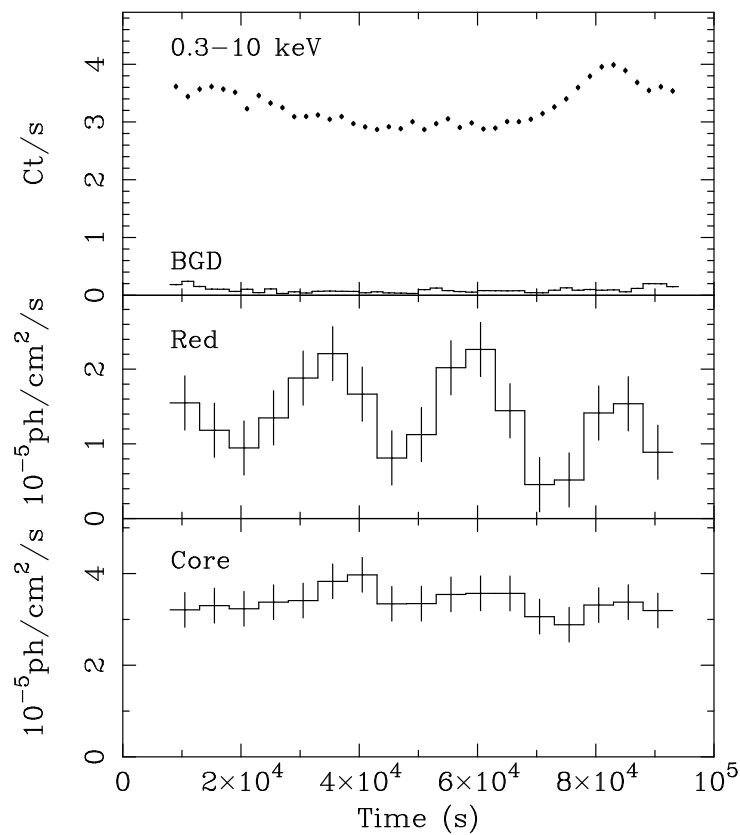}
\caption{Light curves of Seyfert galaxy NGC 3516 \citep{Iwasawa04} for:
the 0.3 -- 10 keV band (top), the Fe K$\alpha$ line red feature (middle) and the 6.4 keV line core (bottom).}
\label{fig61_1}
\end{figure}

In some cases the observed X-ray variability of AGN cannot be
explained by the standard model of accretion disk. For example, in
addition to the stable 6.4 keV core of the Fe K$\alpha$ line, a
variable "red" feature of the line at 6.1 keV is also detected in
X-ray spectrum of Seyfert galaxy NGC 3516 (see Fig. \ref{fig61_1}),
observed by \emph{XMM-Newton} satellite \citep{Iwasawa04}. This
feature varies systematically in the flux at intervals of 25 ks and
in energy between 5.7 and 6.5 keV. \citet{Iwasawa04} found that the
spectral evolution of the "red" feature agrees well with hypothesis
of an orbiting spot in the accretion disk.

Many processes in the accretion disk may lead to perturbations in
its emissivity, such as self gravity, disk-star collisions and
baroclinic vorticity \citep{Flohic08}. Different models of
emissivity perturbing regions can be used to describe the observed
variability in the Fe K$\alpha$ line profiles and intensities, like
stochastically perturbed one given by \citet{Flohic08}, but here we
will present a bright spot model given by \citet{Jovanovic08b}. In
this model a modification of the power-law disk emissivity is
proposed in order to explain the observed profiles. The following
emissivity law of the disk is assumed:
\begin{equation}
\varepsilon_1 (x_p,y_p ) = \varepsilon (r(x_p ,y_p )) \cdot \left( {1 + \varepsilon _p  \cdot e^{ - \left( {\left( {\dfrac{{x - x_p }}{{w_x }}} \right)^2  + \left( {\dfrac{{y - y_p }}{{w_y }}} \right)^2 } \right)} } \right),
\label{eq61_1}
\end{equation}
where $\varepsilon_1 (x_p,y_p )$ is the modified disk emissivity at
the given position $(x_p,y_p)$ of perturbing region (in
gravitational radii $R_g$), $\varepsilon (r(x_p ,y_p ))$ is the
power-law disk emissivity at the same position, $\varepsilon_p$ is
emissivity of perturbing region and $(w_x, w_y)$ are its widths
(also in $R_g$).

\begin{figure}[ht!]
\centering
\vspace*{0.5cm}
\includegraphics[width=\textwidth]{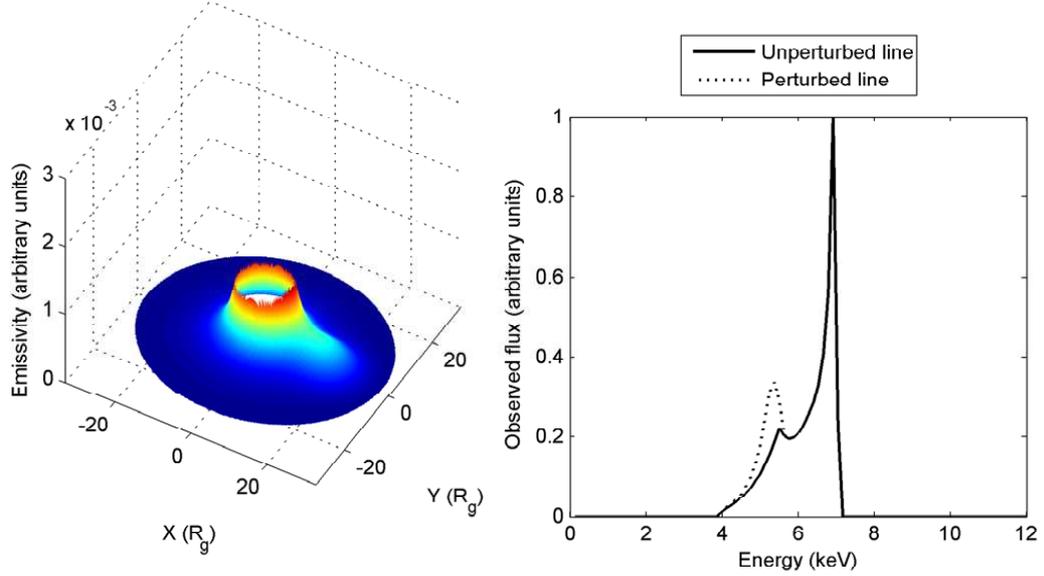}
\caption{\textit{Left:} shape of perturbed emissivity of an accretion
disk in Schwarzschild metric for the following parameters of perturbing region:
$x_p=20\ R_g$, $y_p=0$ and $w_x=w_y=7\ R_g$ \citep{Jovanovic08b}.
\textit{Right:} the corresponding perturbed (dashed line) and
unperturbed (solid line) Fe K$\alpha$ line profiles.}
\label{fig61_2}
\end{figure}

An example of the shape of the perturbed emissivity for an accretion
disk in Schwarzschild metric, as well as the corresponding perturbed
and unperturbed Fe K$\alpha$ line profiles, are presented in the
left and right panels of Fig. \ref{fig61_2}, respectively.

As one can see from Fig. \ref{fig61_2}, when perturbation moves
along receding side of the disk (positive direction of $x$-axis),
the perturbing model affects only "red" part of the line flux, while
the "blue" one stays nearly constant, as well as the line core.
Therefore, this bright spot model of perturbing region could
satisfactorily explain the observed variations of the Fe K$\alpha$
line flux.

Under assumption that perturbation moves by speed of light $c$, one
can also calculate time $t_p\left[s\right]$ that corresponds to the
current position $(x_p,y_p)$ of perturbation, using the following
expression \citep{Jovanovic08b}:
\begin{equation}
t_p \left[ s \right] = \dfrac{{r(x_p ,y_p )\left[ {R_g } \right]}}{{c\left[ {m \cdot s^{ - 1} } \right]}} = \dfrac{{r_{x,y}  \cdot GM_{BH} }}{{c^3 }},
\label{eq61_2}
\end{equation}
where $r_{x,y}  = \dfrac{{r\left( {x_p ,y_p } \right)}}{{R_g }},$
$G$ is Newton's gravitational constant and $M_{BH}$ is the mass of
central black hole.

\begin{figure}[ht!]
\centering
\includegraphics[width=0.85\textwidth]{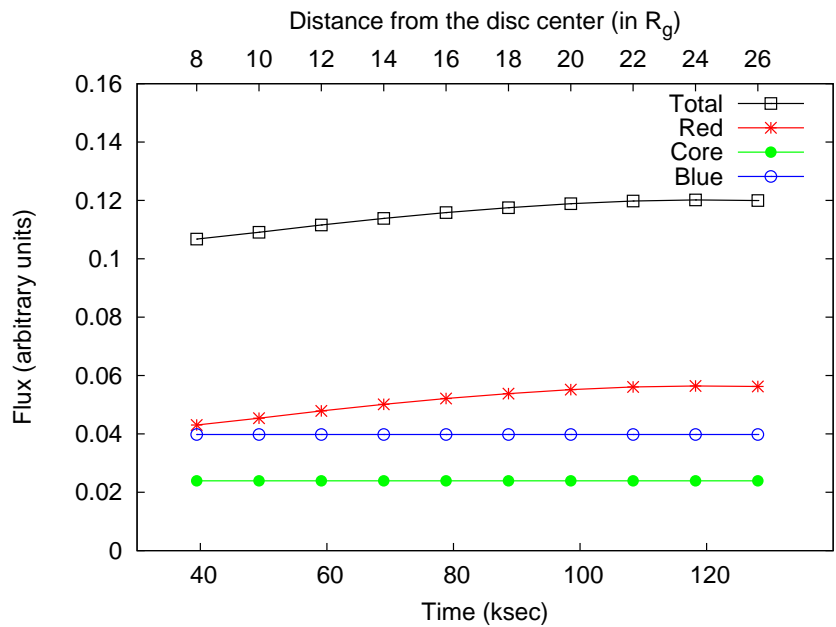}
\caption{The simulated light curves in case of perturbing region displacement along the receding side of the disk and without rotation \citep{Jovanovic08b}.
Light curves correspond to the following spectral bands: total flux (black) to
0.1 -- 12.8 keV,  "red" to 0.1 -- 6.1 keV, "core" to 6.1 -- 6.7 keV and " blue" to 6.7 -- 12.8 keV.}
\label{fig61_3}
\end{figure}

Using the time $t_p\left[s\right]$ we are now able to obtain
simulated light curves, produced as perturbation moves along the
accretion disk. An example of such light curves, corresponding to
displacement of perturbation along the receding side of the disk, is
given in Fig. \ref{fig61_3}. From this figure it can be seen that
displacement of perturbing region results in variations of only
"red" light curve (0.1 -- 6.1 keV), while the "blue" one (6.7 --
12.8 keV) and the line core (6.1 -- 6.7 keV) stay nearly
constant.\footnote{We should note here that we did not take into
account the rotation which can produce periodical peaks in the light
curves, as it can be seen from Fig. \ref{fig61_1}} These variations
are then reflected in total line flux in 0.1 -- 12.8 keV energy band
\citep{Jovanovic08b}.

Thus, this perturbing model could explain the variable "red" feature
of the Fe K$\alpha$ line observed in NGC 3516. Besides, the
realistic durations of disk emissivity perturbations are also
obtained if a central supermassive black hole with mass $M_{BH}=1
\times 10^9 M_\odot$ is assumed \citep{Jovanovic08b}.

\clearpage

\subsection{Absorption by warm X-ray absorbers}

\begin{figure}[b!]
\centering
\includegraphics[width=0.8\textwidth]{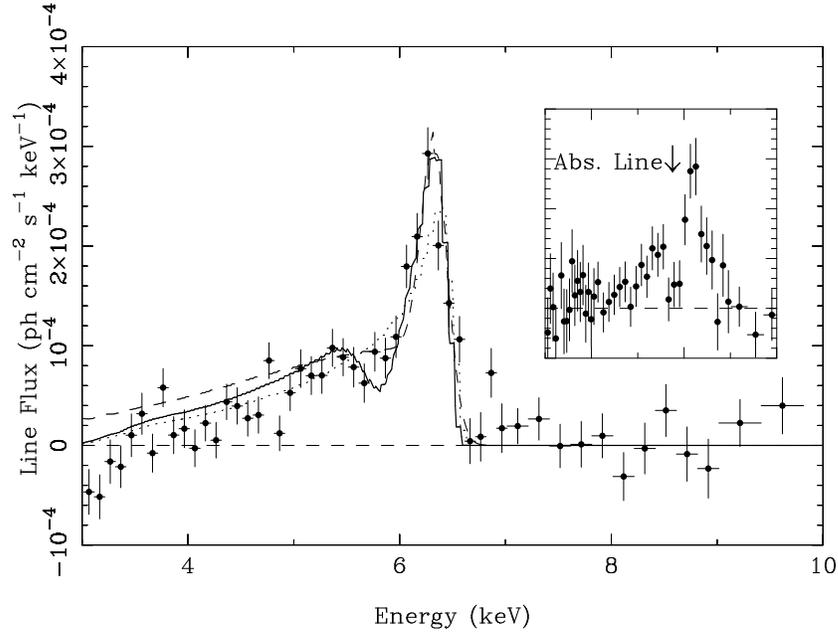}
\caption{Absorption component at 5.8 keV in the red part
of the Fe K$\alpha$ line of Seyfert 1 galaxy NGC 3516, observed by \emph{ASCA} satellite \citep{Nandra99}.
The dashed line shows best fit with model of an accretion disk around a rotating (Kerr) black hole.}
\label{fig62_1}
\end{figure}

The X-ray emission of AGN could be also significantly absorbed by an
outflowing wind, especially in case of so-called Low Ionization
Broad Absorption Line (LoBAL) quasars. Recent observations of such
quasars (e.g. Mrk 231 \citep{Braito04} and H 1413+117
\citep{Chartas07}) confirmed the presence of X-ray absorbers in
these objects. \citet{Wang01} detected an absorption line at 5.8 keV
in nearby ($z=0.0033$) Seyfert 1.5 galaxy NGC 4151. A variable
absorption line at the same energy has been discovered by
\citet{Nandra99} in NGC 3516 (Fig. \ref{fig62_1}) and was
interpreted as a Fe K resonant absorption line, redshifted either by
infalling absorbing material or by strong gravity in the vicinity of
the black hole.

\begin{figure}[ht!]
\centering
\includegraphics[width=0.6\textwidth]{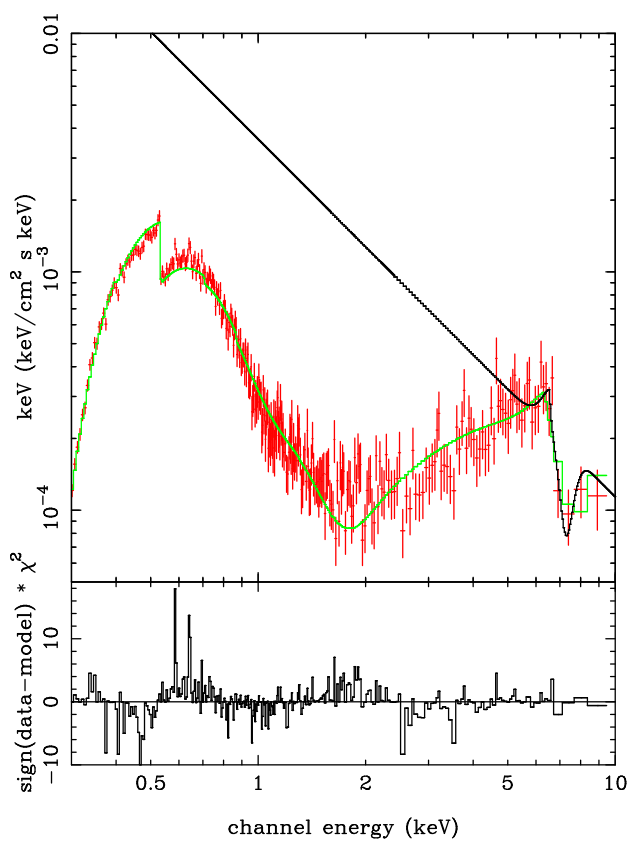}
\caption{The full \emph{XMM-Newton} spectrum of 1H 0707-495 with the
best-fit model which involves a P Cygni profile for the iron
features \citep{Done07}. The lower panel shows residuals to the fit.}
\label{fig62_2}
\end{figure}

\citet{Done07} found an evidence for a P Cygni profile of the Fe
K$\alpha$ line (Fig. \ref{fig62_2}) in narrow line Seyfert 1
galaxies. According to these authors, complex X-ray spectra of these
objects show strong "soft excess" below 2 keV and a sharp drop at
$\sim 7$ keV which can be explained either by reflection or by
absorption from relativistic, partially ionized material close to
the black hole. They showed that a sharp feature at $\sim 7$ keV
results from absorption/scattering/emission of the iron K$\alpha$
line in the wind. In the case of 1H 0707-495 (Fig. \ref{fig62_2}),
this absorption feature can be satisfactorily fitted by the P Cygni
profile \citep{Done07}.

\begin{figure}[hb!]
\centering
\includegraphics[width=\textwidth]{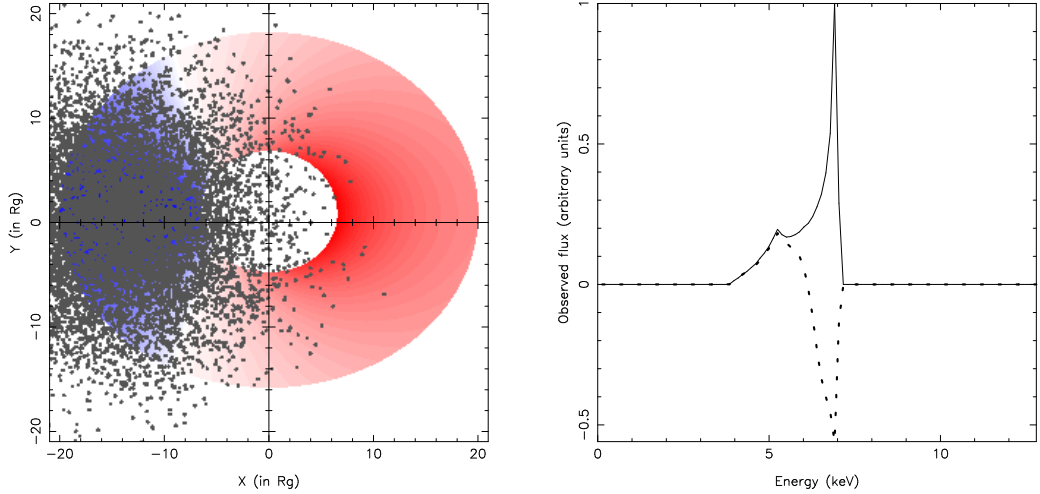}
\caption{\emph{Left:} Relativistic accretion disk in the Schwarzschild metric partially covered
by a cloud of absorbing material (randomly scattered gray dots).
\emph{Right:} Comparison between the unabsorbed Fe K$\alpha$ spectral line profile (solid line)
and corresponding absorbed profile (dotted line) caused by the absorbing/obscuring region presented in the left panel \citep{Jovanovic07}.}
\label{fig62_3}
\end{figure}

There are different models of the X-ray absorbing/obscuring regions,
like absorbing medium comprised of cold absorbing cloudlets by
\citet{Fuerst04}, but here we will focus on the model given by
\citet{Jovanovic07}, developed in order to study how much warm
absorbers can change the Fe K$\alpha$ spectral line profile, emitted
from a relativistic accretion disk. In this model, absorption region
is considered to be composed of a number of individual spherical
absorbing clouds with the same small radii (see Fig. \ref{fig62_3}
left), scattered in space so that projections of their centers to
the observer's sky plane $(X_i, Y_i)$ have bivariate normal
distribution ${\cal N}_2\left(\mu,\Sigma\right)$. Here,
$\mu=\left[\mu_X,\mu_Y\right]^T$ and $\Sigma=\left[
{\begin{array}{*{20}c}
   {\sigma _X^2 } & {\rho \sigma _X \sigma _Y }  \\
   {\rho \sigma _X \sigma _Y } & {\sigma _Y^2 }  \\
\end{array}} \right]$, where $\mu_X$ and $\mu_Y$ are the means of $X_i$ and
$Y_i$, $\sigma _X$ and $\sigma _Y$ are their standard deviations and
$\rho$ is the correlation between them. Absorbing region presented
in Fig. \ref{fig62_3} (left) is obtained for the following
parameters: $\rho=0$, $\mu_X=X_A$, $\mu_Y=Y_A$ and $\sigma_X =
\sigma_Y = R_A$, where $(X_A, Y_A)$ is the center and $R_A$ is the
radius of projection of entire absorbing region.

The absorption coefficient $A(X,Y)$ for every spherical cloud in the
absorbing region is given by \citep{Jovanovic07}:
\begin{equation}
A(X,Y) = \left( 1-I_A(X,Y)\right) \cdot e^{-\left(\dfrac{g(X,Y)\
E_0-E_A}{\sigma_E}\right)^2},
\label{eq62_1}
\end{equation}
where absorption intensity coefficient $I_A(X,Y)$ describes the
distribution of absorption over the whole region, $E_A$ is the
central energy of absorption and $\sigma_E$ is the width of
absorption band (velocity dispersion).

A comparison between the unabsorbed Fe K$\alpha$ spectral line
profile and the corresponding absorbed profile obtained by this
absorption model is given in Fig. \ref{fig62_3} (right). As it can
be seen from Fig. \ref{fig62_3}, when the X-ray radiation from
approaching side of the disk is significantly absorbed/obscured by
the absorbing region, there is a very strong absorption of the iron
line. In such case the emission Fe K$\alpha$ line looks redshifted
at $\sim 5$ keV and is followed by a strong absorption line at $\sim
7$ keV (Fig. \ref{fig62_3} right), which indicates the P Cygni
profile of the iron line. Thus, this model can satisfactorily
explain the P Cygni profile of the Fe K$\alpha$ line in the case
when approaching side of the accretion disk is partially blocked
from our view by the X-ray absorbing/obscuring material, while the
rest of the disk is less absorbed/obscured and therefore is visible
\citep{Jovanovic07}.

\subsection{Gravitational microlensing}

Some recent observational and theoretical studies suggest that
gravitational microlensing can also induce variability in the X-ray
emission of AGN, especially in the case of gravitationally lensed
quasars. The phenomenology of gravitational lensing effects and
introduction to this field has been given in several review papers
and books. References to some of them can be found in e.g.
\citet{Popovic02}. Therefore, in this section we will not discuss
all aspects of such effects in the universe, but instead, we will
briefly present the basic concepts of gravitational microlensing
theory, as well as some examples for microlensing influence on the
X-ray emission from AGN.

\begin{figure}[ht!]
\centering
\includegraphics[width=\textwidth]{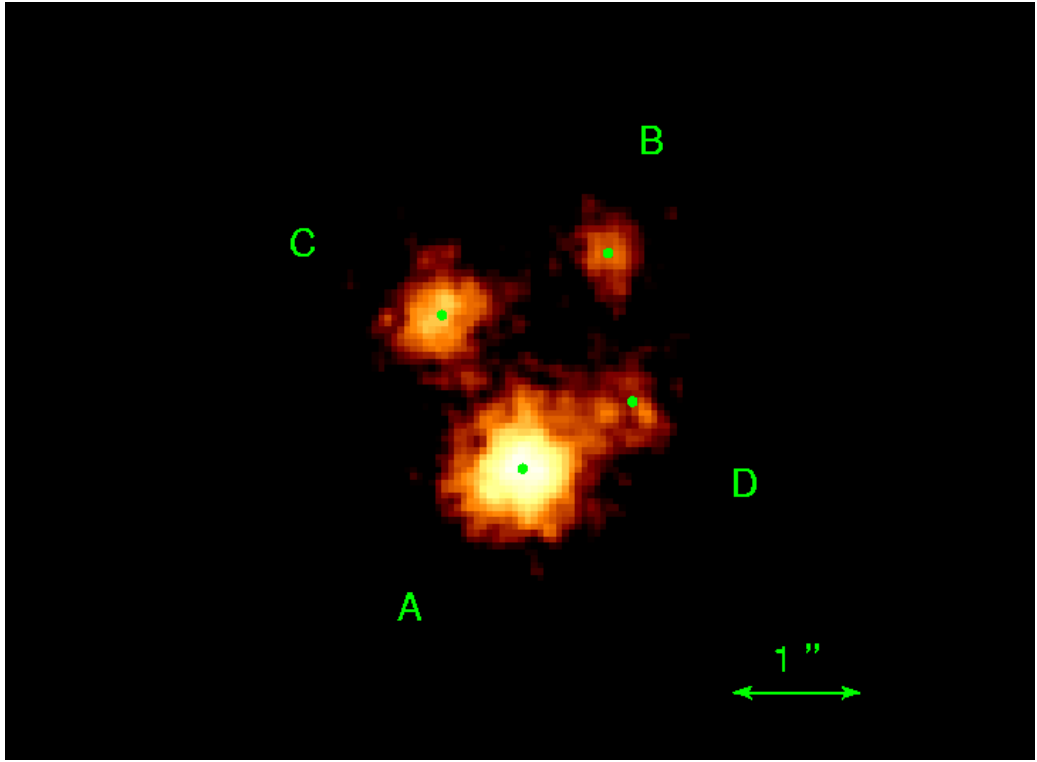}
\caption{Gravitationally lensed system QSO 2237+0305 (Einstein Cross)
observed by the \emph{Advanced CCD Imaging Spectrometer (ACIS)}
onboard the \emph{Chandra X-ray Observatory} \citep{Dai03}. The green circles in
each image are the corresponding \emph{Hubble Space Telescope (HST)}
image positions provided by \emph{CASTLES} (http://cfawww.harvard.edu/glensdata/Individual/Q2237.html).}
\label{fig63_0}
\end{figure}

Gravitational lensing is an universal natural phenomenon where the
gravitational force of lensing object induce either the
amplification of some background source (\textbf{microlensing}), or
the appearance of its multiple images (\textbf{macrolensing}), due
to light bending in a gravitational field of the deflector.
Separation angle between the images of background source depends on
the mass of gravitational lens and therefore, multiple images can be
observed only in the case of a massive lensing object, such as
galaxy. One of the most famous multiple image lens systems is quasar
QSO 2237+0305, also known as Einstein Cross, which is located at
redshift $z=1.695$ (see Fig. \ref{fig63_0}). Its four images are due
to lensing effect of galaxy ZW2237+030, located between us and the
quasar at redshift $z=0.0394$. In the case of a small mass lens
(e.g. a star), the separation angle is also small and therefore,
different images of background source cannot be resolved. Instead,
its intensity is amplified, causing the changes in the observed
light curve. Common name for both, macrolensing (or simply lensing)
and microlensing is \textbf{strong lensing}, contrary to
\textbf{weak lensing}, which causes distortions in observed images
of distant objects which can be then used for studying the mass
distribution along the line of sight, but such phenomena will not be
discussed here.

\citet{Zakharov04} found that cosmologically distributed microlenses
could significantly contribute to the X-ray variability of
high-redshifted ($z>2$) QSOs. Indeed, microlensing of the Fe
K$\alpha$ line has been reported at least in three macro\-lensed
QSOs: MG J0414+0534 \citep{Chartas02}, QSO 2237+0305 \citep{Dai03},
and H 1413+117 \citep{Oshima01,Chartas04}. \citet{Chartas02}
detected an increase of the Fe K$\alpha$ equivalent width in the
image B of MG J0414+0534 which was not followed by the continuum and
explained this behavior by assumption that the thermal emission
region of the disk and the Compton up-scattered emission region of
the hard X-ray source lie within smaller radii than the iron-line
reprocessing region. Analyzing the X-ray variability of QSO
2237+0305A, \citet{Dai03} also measured amplification of the Fe
K$\alpha$ line in component A of QSO 2237+0305 but not in the
continuum and suggested that the larger size of the continuum
emission region in comparison to the Fe K$\alpha$ emission region
could explain this result. In the case of H 1413+117,
\citet{Chartas04} found that the continuum and the Fe K$\alpha$ line
were enhanced by a different factor. For more detailed discussion
about observational and theoretical investigations of the X-ray
variability in gravitationally lensed QSOs see e.g.
\citet{Jovanovic03,Jovanovic05,Jovanovic05a,Jovanovic06,Jovanovic08}
and \citet{Popovic06a}, and also references therein.

The influence of gravitational microlensing on the X-ray emission
from a compact accretion disk of AGN can be studied using numerical
simulations based on ray-tracing method described in \S 5.3 of this
chapter. In the following text we will briefly review three models
of gravitational microlenses (point-like, straight-fold caustic and
quadruple microlens), as well as their influence on the X-ray
radiation of AGN.

If $X$ and $Y$ are the impact parameters (i.e. the coordinates)
which describe the apparent position of each point of the accretion
disk image on the celestial sphere as seen by an observer at
infinity, then the amplified brightness (due to gravitational
microlensing influence) is given by
\citep{Jovanovic03,Popovic03a,Popovic03b}:
\begin{equation}
I_p(X,Y)=\varepsilon(X,Y)g^4(X,Y)\delta(x-g(X,Y))A(X,Y),
\label{eq63_1}
\end{equation}
where $x=\nu_{obs}/\nu_0$ ($\nu_0$ and $\nu_{obs}$ are the
transition and observed frequencies, respectively),
$g=\nu_{obs}/\nu_{em}$ is the shift due to relativistic effects
($\nu_{em}$ is the emitted frequency), $\varepsilon$ is the disk
emissivity and $A$ is the amplification caused by microlensing.
Usually, we do not know the exact form of $A$, so we are forced to
consider the different approximations in order to estimate this
quantity.

The first of these approximations is called \textbf{point-like
microlens}, and it is applied when an isolated compact object (e.g.
a star) plays a role of gravitational microlens. Such microlens is
characterized by its Einstein Ring Radius in the lens plane
\citep{Jovanovic08}:
\begin{equation}
ERR=\sqrt{\dfrac{4Gm}{c^2}\dfrac{D_lD_{ls}}{D_s}}
\end{equation}
or by the corresponding projection to the source plane:
\begin{equation}
\eta_0=\dfrac{D_s}{D_l}ERR=\sqrt{\dfrac{4Gm}{c^2}\dfrac{D_sD_{ls}}{D_l}},
\end{equation}
where $G$ is the gravitational constant, $c$ is the speed of light,
$m$ is the microlens mass and $D_l$, $D_s$ and $D_{ls}$ are the
cosmological angular distances between observer-lens,
observer-source and lens-source, respectively. In this case, the
amplification is given by the following relation
\citep{Narayan99,Popovic01,Jovanovic03}:
\begin{equation}
A(X,Y)=\frac{u^2(X,Y)+2}{u(X,Y)\sqrt{u^2(X,Y)+4}},
\label{eq63_2}
\end{equation}
where
\begin{equation}
u(X,Y)=\dfrac{\sqrt{(X-X_0)^2 +(Y-Y_0)^2}}{\eta_0}
\end{equation}
corresponds to the angular separation between the microlens and a
source. $X_0,Y_0$ are the coordinates of the microlens with respect
to the disk center (given in $R_g$).

The total observed flux is then given by
\citep{Jovanovic03,Popovic03a,Popovic03b}:
\begin{equation}
F(x)=\int_{image}I_p(x)d\Omega
\label{eq63_3}
\end{equation}
 where $d\Omega$ is the
solid angle subtended by the disk in the observer's sky and the
integral extends over the whole disk image.

\begin{figure}[ht!]
\centering
\includegraphics[width=0.75\textwidth]{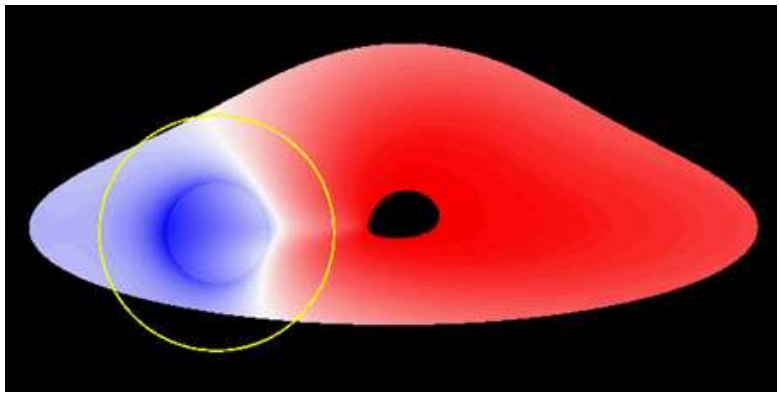}
\caption{Illustration of a point-like gravitational microlens crossing over an accretion disk in Kerr metric
with angular momentum $a=0.998$ \citep{Popovic02,Popovic03b}.
Einstein Ring of the microlens is schematically presented by yellow Euclidian circle.}
\label{fig63_1}
\end{figure}

\begin{figure}
\centering
\includegraphics[width=\textwidth]{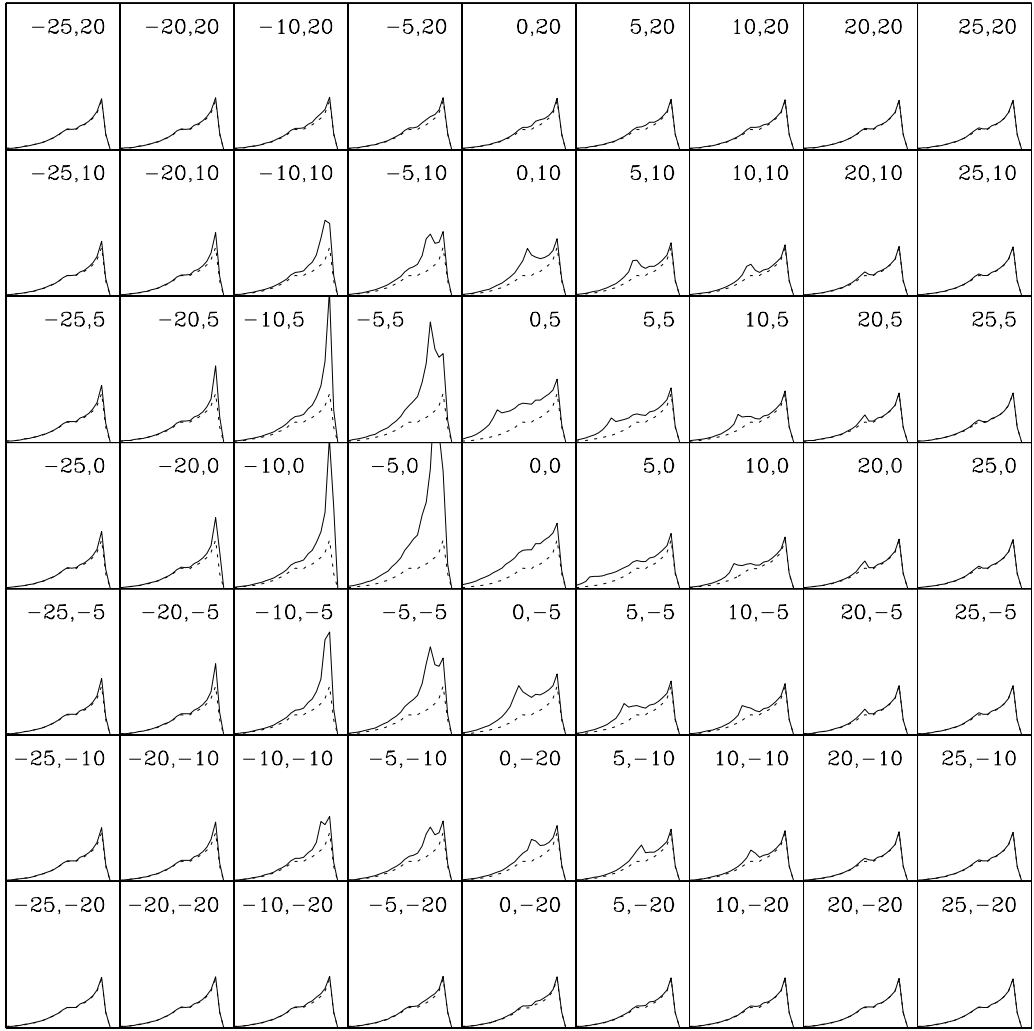}
\caption{The unamplified (dotted line) and amplified Fe K$\alpha$ line profiles (solid line) for different positions of the point-like microlens, which
coordinates (expressed in $R_g$) in respect to the accretion disk center are given at the top of each panel \citep{Popovic03a}.
The simulations are performed for an accretion disk in Kerr metric with angular momentum $a=0.998$ and
with the following disk parameters: R$_{in}=6\ R_g$, R$_{out}=20\
R_g$, $i=35^\circ$, q=2.5. The Einstein Ring Radius of the point-like microlens is 10 R$_g$.
The relative intensity ($y$-axis) is in the range from 0 to 3, and energy shift $g$ due to relativistic effects ($x$-axis) is in
the range from 0.4 to 1.2.}
\label{fig63_2}
\end{figure}

An illustration of a point-like gravitational microlens crossing
over an accretion disk in the Kerr metric with angular momentum
$a=0.998$ is presented in Fig. \ref{fig63_1}
\citep{Popovic02,Popovic03b}. The simulated unamplified and
amplified Fe K$\alpha$ line profiles due to different positions of
the microlens are shown in Fig. \ref{fig63_2} \citep{Popovic03a}.

From Fig. \ref{fig63_2} it can be seen that point-like microlens
could induce strong changes of the Fe K$\alpha$ line shape and
intensity, depending on the location of the microlens. In the first
place, we have in mind the changes in number of peaks, their
relative separation and the peak velocity. Secondly, such transit of
the point-like microlens could cause an asymmetrical enhancement of
the line profile, and the maximum of the amplification would be
obtained for negative values of $x$-coordinate, i.e. for the
approaching side of the rotating accretion disk, and therefore it
would affect mainly the blue part of the Fe K$\alpha$ line.

In most cases we cannot simply consider that microlensing is caused
by an isolated compact object, but we should take into account that
such micro-deflector is located in an extended object (typically,
the lens galaxy). Therefore, when the size of the microlens
projected Einstein Ring Radius is larger than the size of the
accretion disk and when a number of microlenses form a caustic net,
we describe the microlensing effects by a \textbf{straight-fold
caustic} approximation \citep{Jovanovic08,Popovic03a}, where the
amplification at a point of an extended source (accretion disk)
close to the caustic is given by \citep{Chang84}:
\begin{equation}
A(X,Y)=A_0+K\sqrt{\frac{r_{caustic}}{\kappa(\xi-\xi_c)}}\cdot H(\kappa(\xi-\xi_c)).
\label{eq63_4}
\end{equation}
In above expression $A_0$ is the amplification outside the caustic
and $K=A_0\beta$ is the caustic amplification factor, where $\beta$
is constant of order of unity. $\xi$ is the distance perpendicular
to the caustic and $\xi_c$ is the minimum distance from the disk
center to the caustic (both in $R_g$). The "caustic size"
$r_{caustic}$ is the distance $\xi$ for which the caustic
amplification is equal to 1, and therefore this parameter defines a
typical linear scale for the caustic. $H(\kappa(\xi-\xi_c))$ is the
Heaviside function which equals 1 for $\kappa(\xi-\xi_c)>0$ and
otherwise it is 0. $\kappa$ is $\pm 1$ depending on the direction of
caustic motion: if the direction of the caustic motion is from
approaching side of the disk $\kappa=-1$, otherwise it is +1. Also,
in the special case of caustic crossing perpendicular to the
rotating axis $\kappa=+1$ for direction of caustic motion from $-Y$
to $+Y$, otherwise it is $-1$.

A microlensing event where a caustic crosses over an accretion disk
can be described in the following way \citep{Jovanovic08}: before
the caustic reaches the disk, the amplification is equal to $A_0$
because the Heaviside function of Eq. (\ref{eq63_4}) is zero. Just
as the caustic begins to cross the disk, the amplification rises
rapidly and then decays gradually towards $A_0$ as the source moves
away from the caustic-fold.

\begin{figure}[ht!]
\centering
\includegraphics[width=0.75\textwidth]{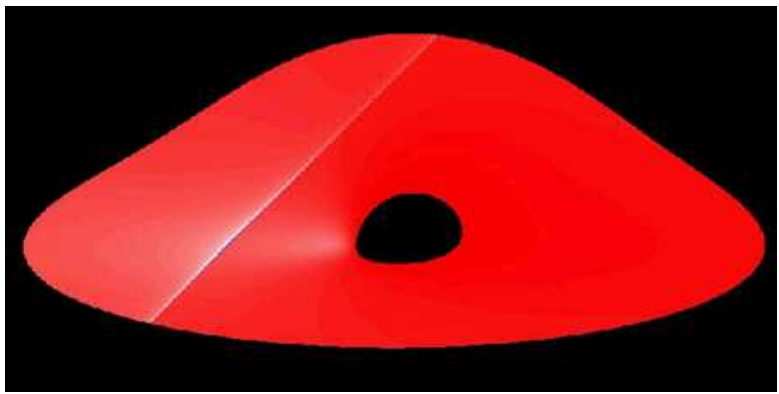}
\caption{Illustration of a straight-fold caustic gravitational microlens crossing over an accretion disk in Schwarzschild metric \citep{Jovanovic06,Popovic03b}.\label{fig63_3}}
\end{figure}

\begin{figure}
\centering
\includegraphics[width=\textwidth]{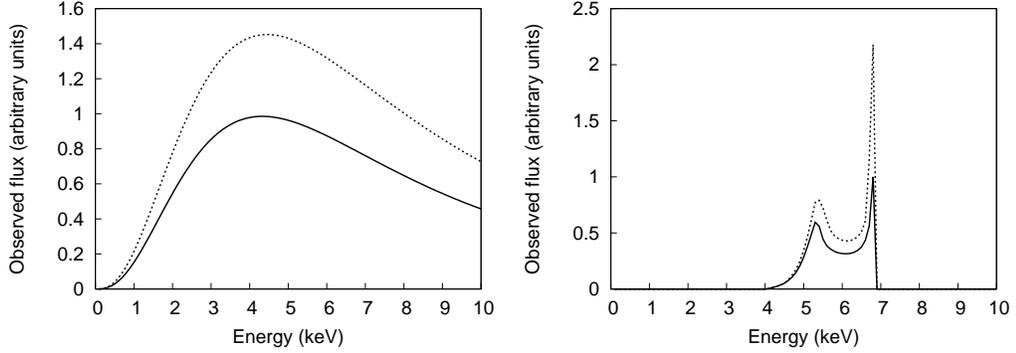}
\caption{The deformations of the X-ray continuum (left) and the Fe K$\alpha$ line (right) due to influence of straight-fold caustic from Fig. \ref{fig63_3}.
Undeformed profiles are presented by solid and deformed by dotted lines \citep{Jovanovic06}.\label{fig63_4}}
\end{figure}

An illustration of the straight-fold caustic crossing over an
accretion disk in Schwarzschild metric is presented in Fig.
\ref{fig63_3}, and the corresponding effects on the shapes of the
X-ray continuum and the Fe K$\alpha$ line are shown in Fig.
\ref{fig63_4} \citep{Jovanovic06,Popovic03b}. As it can be seen from
Fig. \ref{fig63_4}, caustics could also induce significant
amplifications of both, the X-ray continuum and Fe K$\alpha$ line.

The most complex approximation for gravitational microlensing
amplification is so called \textbf{quadruple microlens}. This model
is applied to obtain a spatial distribution of magnifications in the
source plane (where an accretion disk of AGN is located), produced
by a random star field placed in the lens plane. Such spatial
distribution of magnifications is called \textbf{microlensing map,
microlensing pattern or caustic network}. If we consider a set of
$N$ compact objects (e.g. stars) which are characterized by their
positions $x_i$ and their masses $m_i$, then the normalized lens
equation is given by \citep{Abajas07}:
\begin{equation}
\vec y = \sum\limits_{i = 1}^N {m_i \frac{{\vec x - \vec x_i }}{{\left| {\vec x - \vec x_i } \right|^2 }}}
+ \left[ {\begin{array}{*{20}c} {1 - \kappa_c  + \gamma } & 0  \\ 0 & {1 - \kappa_c  - \gamma }  \\ \end{array}} \right]\vec x,
\label{eq63_5}
\end{equation}
where $\vec x$ and $\vec y$ are the normalized image and source
positions, respectively. The sum describes the light deflection by
the stars and the last term is a quadruple contribution from the
galaxy containing the stars, where $\kappa_c$ is a smooth surface
mass density and $\gamma$ is an external shear. The total surface
mass density or convergence can be written as
$\kappa=\kappa_\ast+\kappa_c$, where $\kappa_\ast$ represents the
contribution from the compact microlenses. The corresponding
microlensing map is then defined by two parameters: the convergence
- $\kappa$, and the shear due to the external mass - $\gamma$.

\begin{figure}[ht!]
\centering
\includegraphics[width=\textwidth]{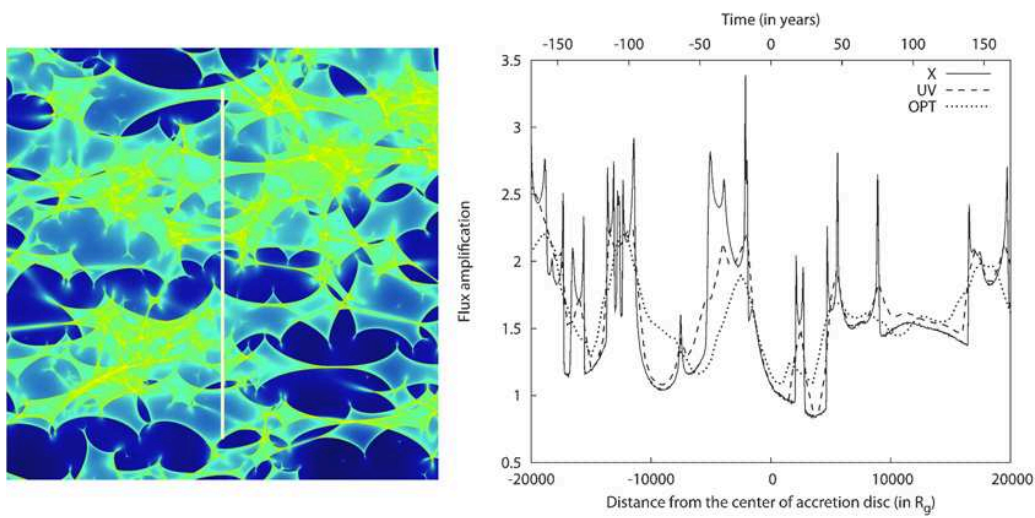}
\caption{\emph{Left:} Magnification map of a "typical" lens system, where
the white solid line represents a path of an accretion disk center.
\emph{Right:} Variations in the X-ray (solid), UV (dashed) and
optical (dotted) spectral bands corresponding to the path in the magnification map of a "typical" lens system (left).}
\label{fig63_5}
\end{figure}

For some specific lensing event one can model the corresponding
magnification map using numerical simulations based on ray-shooting
techniques (see e.g. \citet{Jovanovic08,Popovic06a,Popovic06b} and
references therein), in which the rays are shot from the observer to
the source, through the randomly generated star field in the lens
plane. An example of a magnification map for a "typical" lens system
(i.e. for a lens system where the redshifts of microlens and source
are $z_l = 0.5$ and $z_s = 2$, respectively) is presented in the
left panel of Fig. \ref{fig63_5}, and the corresponding X-ray, UV
and optical continuum variations are given in the right panel of the
same figure \citep{Jovanovic08}. The size of this magnification map
is $16\ \eta_0\times16\ \eta_0$ and it is generated using the
following arbitrary values for convergence and shear: $\kappa=0.45$
and $\gamma=0.3$. The light curves in the right panel of Fig.
\ref{fig63_5} are produced when an accretion disk crosses over the
magnification pattern in the left panel, along the path denoted by
the white solid line.

As one can see from Fig. \ref{fig63_5} (right), the variations of
the X-ray continuum due to microlensing are much stronger and faster
in comparison to the variations in UV and optical spectral bands
\citep{Jovanovic08,Popovic06a,Popovic06b}. Usually, so called High
Amplification Events, i.e. asymmetric peaks in the light curves are
also analyzed \citep{Jovanovic08}. The results presented in Fig.
\ref{fig63_5} (right) show that the rise times of such events are
the shortest and their frequency the highest in the X-rays, in
comparison to the UV/optical spectral bands.

\begin{figure}[ht!]
\centering
\includegraphics[width=0.75\textwidth]{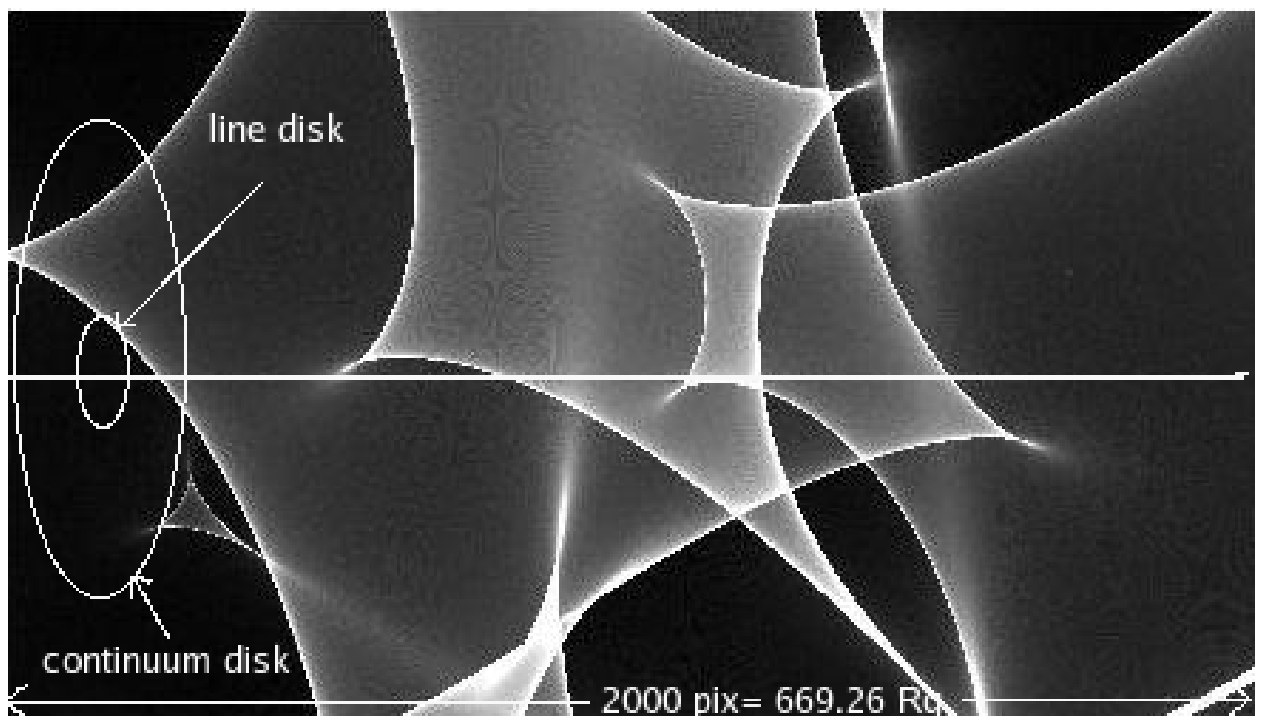} \\
\includegraphics[width=\textwidth]{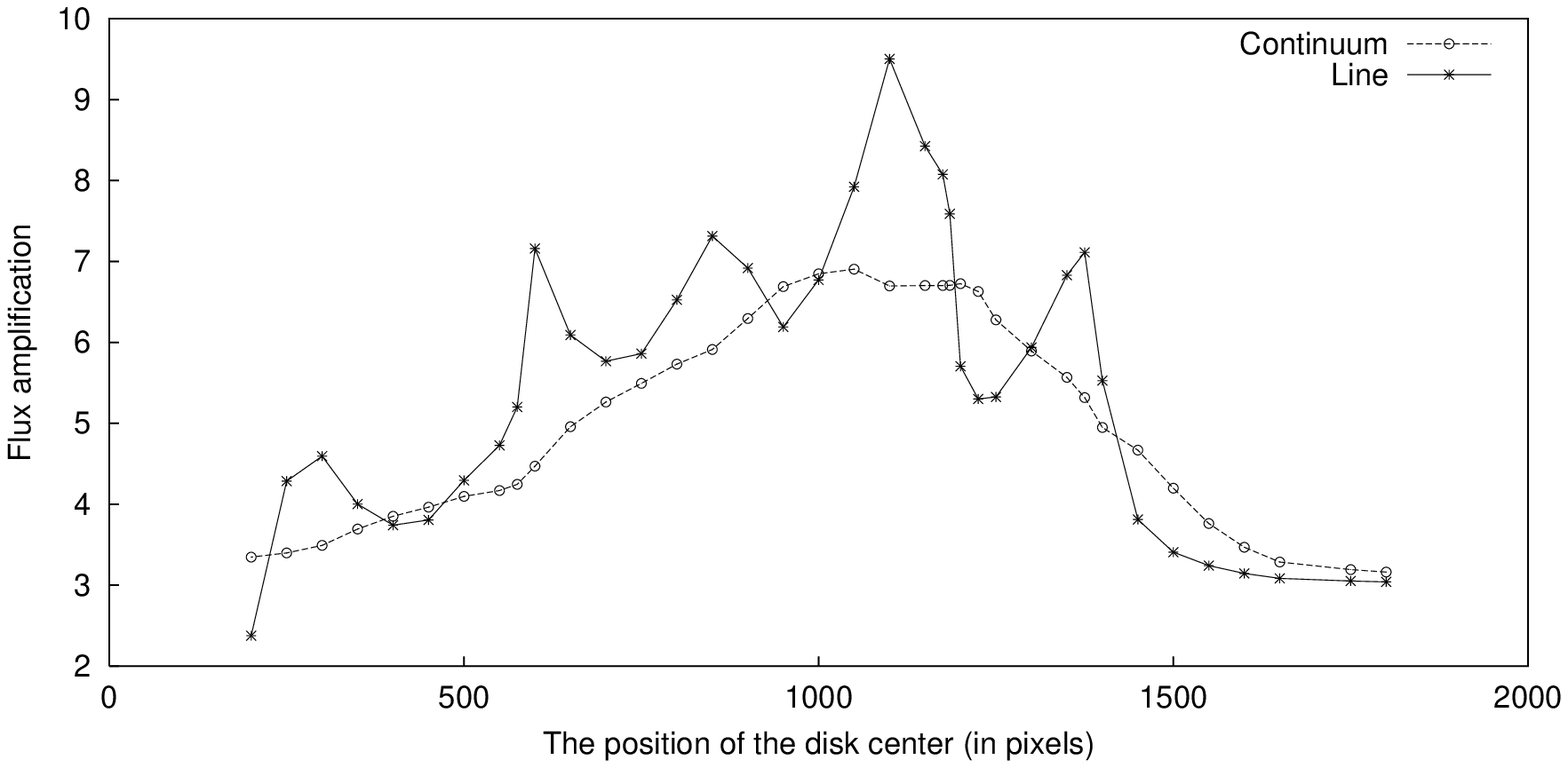}
\caption{\emph{Top:} Microlensing magnification map of QSO 2237+0305A image with $1\ \eta_0\times 2\ \eta_0$
on a side and a scheme of an accretion disk with separated Fe K$\alpha$ line and X-ray continuum emitting regions
which outer radii are 20 and 100 R$_g$, respectively \citep{Popovic06a}. The straight horizontal line represents a path of the disk center.
\emph{Bottom:} The corresponding amplifications of the Fe K$\alpha$ line and X-ray
continuum total fluxes for different positions of the disk center along the path on the magnification map in the left panel.}
\label{fig63_6}
\end{figure}

In order to explain the observed X-ray variability in
gravitationally lensed quasars, \citet{Popovic06a} considered a
microlensing magnification map for a specific case of the QSO
2237+0305A image (Fig. \ref{fig63_6}, top). This map with $1\
\eta_0\times 2\ \eta_0$ on a side was generated using the following
parameters: $\kappa=0.36$, $\gamma= 0.40$ and the mean mass of
randomly distributed deflectors $\langle m\rangle=0.35\ M_\odot$.
The corresponding variations of the total Fe K$\alpha$ line and
X-ray continuum fluxes are presented in Fig. \ref{fig63_6} (bottom).

As one can see from Fig. \ref{fig63_6} (bottom), there is a global
correlation between the total line and continuum fluxes during the
complete simulated microlensing event, but while the continuum flux
variations are smooth and slow, the line flux varies much stronger
and faster \citep{Jovanovic05}. Moreover, there is a certain period
in the middle of the microlensing event in which the Fe K$\alpha$
line flux changes rapidly, while the continuum flux remains nearly
constant. This indicates that different behavior of the line and
X-ray continuum variability, observed in some lensed quasars
\citep{Chartas02,Chartas04}, may be explained by microlensing
hypothesis if the line originates from the innermost part of the
accretion disk, while the X-ray continuum is emitted from a larger
region in the disk
\citep{Jovanovic03,Jovanovic05,Jovanovic05a,Popovic06a}.

On the basis of all facts presented in this section, one can
conclude that gravitational microlensing can produce significant
variations and amplifications of the Fe K$\alpha$ line and X-ray
continuum. During a microlensing event, even very small mass objects
could produce noticeable changes in the X-ray radiation from
accretion disks of AGN. These changes are significantly larger than
the corresponding effects in the optical and UV emission. Also,
microlensing hypothesis can satisfactorily explain the excess in the
iron line emission, observed in some gravitational lens systems.

\section{Conclusion}
\markright{Conclusion}

In this chapter we discussed the X-ray radiation from relativistic
accretion disks around supermassive black holes, supposed to exist
in the centers of all Active Galactic Nuclei. The X-ray radiation is
created when the accreting material loses its angular momentum due
to friction between the adjacent layers in the disk and spiral
towards the central black hole, releasing the gravitational energy.
Part of this energy increases the kinetic energy of rotation and the
rest is turned into the radiation, which is then emitted from the
disk surface in the X-ray, UV and optical spectral bands. It is the
only known mechanism of converting gravitational potential energy
into radiation, which is sufficiently efficient to explain the high
luminosities of some observed AGN.

The standard model of an accretion disk, including its emission,
accretion rate, luminosity, structure and spectral distribution was
described in details. A significant part of this chapter was devoted
to modeling of the observed X-ray radiation in both the Fe K$\alpha$
spectral line and X-ray continuum. The modeled X-ray radiation is
then used to analyze the several effects of strong gravity in the
vicinity of black holes which have the influence on the shapes and
intensities of the observed Fe K$\alpha$ line and X-ray continum. We
showed that the X-ray emission strongly depends on the angular
momentum and mass of the central black hole, as well as on several
parameters describing the X-ray emitting region in the disk. This
was demonstrated on several examples, showing various aspects of the
X-ray emission from the accretion disk around Schwarzschild and Kerr
black holes. Since the X-ray radiation is probably produced in a
very compact region near the black holes, such comparison between
observed and modeled X-ray radiation can provide us some essential
information about the plasma conditions and the space-time geometry
in the vicinity of those black holes.

We also studied some possible causes for the rapid and irregular
variability of the Fe K$\alpha$ line and X-ray continuum, observed
in the X-ray spectra of some AGN. Our results show that, at least,
three mechanisms could explain such variability: perturbations of
disk emissivity, absorption by warm X-ray absorbers and
gravitational microlensing.

Taking all this into account, we can conclude that the X-ray
emission from accretion disks of AGN can be used as a powerful tool
for revealing the physics and geometry in the vicinity of their
central supermassive black holes. But, to use the Fe K$\alpha$ line
profile for such investigations, one should also take into account
the effects which can disturb its profile (such as e.g. absorption,
perturbations in the disk and gravitational lensing).

\vfil\noindent\textit{Acknowledgements.} This work is a part of the
project (146002) "Astrophysical Spectroscopy of Extragalactic
Objects" supported by the Ministry of Science of Serbia.

\clearpage
\markright{References}

\end{document}